\documentclass[letterpaper,english,preprint,aps,prb]{revtex4-2}
\usepackage[utf8]{inputenc}
\usepackage[T1]{fontenc}
\usepackage[english]{babel}
\usepackage{xcolor}
\usepackage{float}
\usepackage{bbding}
\usepackage{amsmath}
\usepackage{amssymb}
\usepackage{graphicx}
\usepackage{bm}
\usepackage{svg}
\usepackage{booktabs}
\makeatletter

\begin{document}

\title{Propulsion of magnetic domain walls via elastic damping in magnetostrictive materials}

\author{Miguel Moreno, Rocio Yanes, Luis Lopez-Diaz}
\affiliation{Departamento de Fisica Aplicada, Universidad de Salamanca. 37008, Salamanca, Spain}
\affiliation{Unidad de Excelencia en Luz y Materia Estructuradas (LUMES), Universidad de Salamanca. Salamanca, Spain}

\date{\today}

\begin{abstract}

The interaction between magnetic and elastic degrees of freedom is increasingly central to advanced spintronics technology. In magnetostrictive materials, a moving domain wall carries a localized elastic deformation. Here, we investigate these elastic deformations and their back-action on field-driven domain wall dynamics using a fully coupled micromagnetic-elastodynamic framework and an analytical collective coordinates model. We show that mechanical damping acts as a non-magnetic dissipation channel, extracting energy from the magnetic texture and sustaining translational domain wall motion even in the complete absence of intrinsic Gilbert damping ($\alpha=0$). Specifically, elastic damping impacts domain wall dynamics in a manner equivalent to magnetic losses, allowing us to map it onto an effective magnetoelastic contribution to the Gilbert damping parameter, $\alpha_\text{me}$. We calculate the localized stress tensor profiles co-moving with the domain wall and show that mechanical dissipation breaks their spatial symmetry. Using our collective coordinates model, we demonstrate that these asymmetric stress profiles—specifically the localized stress gradients across the wall core—exert direct propelling forces that balance field-induced magnetic precession, enabling steady-state motion. Overall, these findings reveal a fundamental energy transfer and relaxation mechanism between magnetic and elastic subsystems, providing valuable insights for controlling magnetic texture dynamics via strain engineering and mechanical damping design in acoustic spintronic devices.  

\end{abstract}

\maketitle

\section{Introduction}
\label{sec_intro}

Acoustic spintronics has emerged as a promising paradigm for controlling magnetization dynamics, offering high energy efficiency due to the absence of Joule heating. By exploiting magnetoelastic coupling, mechanical strain provides a versatile means to manipulate magnetic states without relying on high electric currents. Significant progress has been made in this direction, including elastically driven ferromagnetic resonance using surface acoustic waves (SAWs) \cite{Weiler_11,Dreher_12}, reversible and non-volatile magnetization reorientation via localized strain \cite{Wu_11,Ahmad_15}, and acoustic spin current generation alongside spin pumping \cite{Uchida_11,Weiler_12}.

Beyond uniform state manipulation, strain engineering offers high controllability over non-trivial magnetic textures at the nanoscale. Strain gradients and elastic waves have been successfully employed to drive magnetic domain wall (DW) motion \cite{Lei_13,Edrington_18,Rivelles_25}, nucleate and transport topological textures such as skyrmions \cite{Yokouchi_20,Chen_23,Liu_23,Shuai_24}, and resonantly excite the gyrotropic modes of magnetic vortices \cite{Koujok_23}.

When these phenomena are investigated theoretically or computationally, magnetoelastic coupling is most commonly treated as an additional term in the magnetic energy density \cite{Bryan_12,Fattouhi_22,Shuai_24,Fattouhi_25}. In such unidirectional models, the strain field is introduced as a prescribed external perturbation that drives the Landau–Lifshitz–Gilbert (LLG) equation without being affected in return. While these approaches capture key features, they implicitly assume weak coupling and neglect the dynamical back-action exerted by moving magnetization profiles onto the elastic continuum, which can become significant at reduced dimensions.

To account for this mutual interaction, fully coupled models that simultaneously solve elastodynamics and micromagnetics have been developed \cite{Liang_14,Vanderveken_21,Flauger_26}. This bidirectional framework has provided valuable insights into strain-assisted switching \cite{Liang_14}, spin wave interactions with acoustic waves \cite{Babu_21}, magnetoelastic wave propagation \cite{Chen_17,Vanderveken_21,Flauger_26}, topological edge magnons controlled by chiral elastic waves \cite{Yao_26}, magnetoelastic SAW transmission across domain walls \cite{Zang_26}, etc.

In this work, we employ a fully coupled model to investigate field-driven DW motion in a magnetostrictive nanostripe in the absence of external strain. Early theoretical studies on magnetoelastic DW dynamics primarily examined regimes where wall velocities approached or exceeded the sound velocity, leading to resonant Cherenkov-like radiation and supersonic solitary waves \cite{Baryakhtar_78,Zvezdin_92}. In contrast, we focus on the subsonic regime ($v \ll v_s$), isolating how the intrinsic elastic deformations induced by the moving DW itself dictate its dynamical response.

Using micromagnetic-elastodynamic simulations alongside an analytical 1D collective coordinates framework, we reveal how magnetoelastic feedback governs DW propulsion. The remainder of this paper is organized as follows: In Sec.~\ref{sec_model}, we introduce the coupled elastomagnetic continuum formulation and derive the equilibrium magnetoelastic state. Section~\ref{sec_eta_0} examines DW dynamics in the absence of mechanical dissipation ($\eta = 0$), demonstrating that the elastic deformation follows the wall adiabatically and acts as an effective transverse anisotropy. In Sec.~\ref{sec_eta_not_0}, we consider finite mechanical damping ($\eta \neq 0$) and show that elastic dissipation opens a distinct energy relaxation channel capable of driving sustained DW displacement under an applied field, even when intrinsic magnetic Gilbert damping is absent ($\alpha = 0$). To ensure clarity, both Secs.~\ref{sec_eta_0} and \ref{sec_eta_not_0} follow a systematic three-part structure: we first present the full coupled micromagnetic-elastodynamic simulation results, then formulate a tailored collective coordinates model based on co-moving elastic strain, and finally utilize this model to provide a microscopic physical interpretation of the dynamics. Lastly, Sec.~\ref{sec_conclusions} summarizes our main conclusions and outlook.

\section{Model and equilibrium solution}
\label{sec_model}

\begin{figure}[!htb]
    \includegraphics[width=1.0\linewidth]{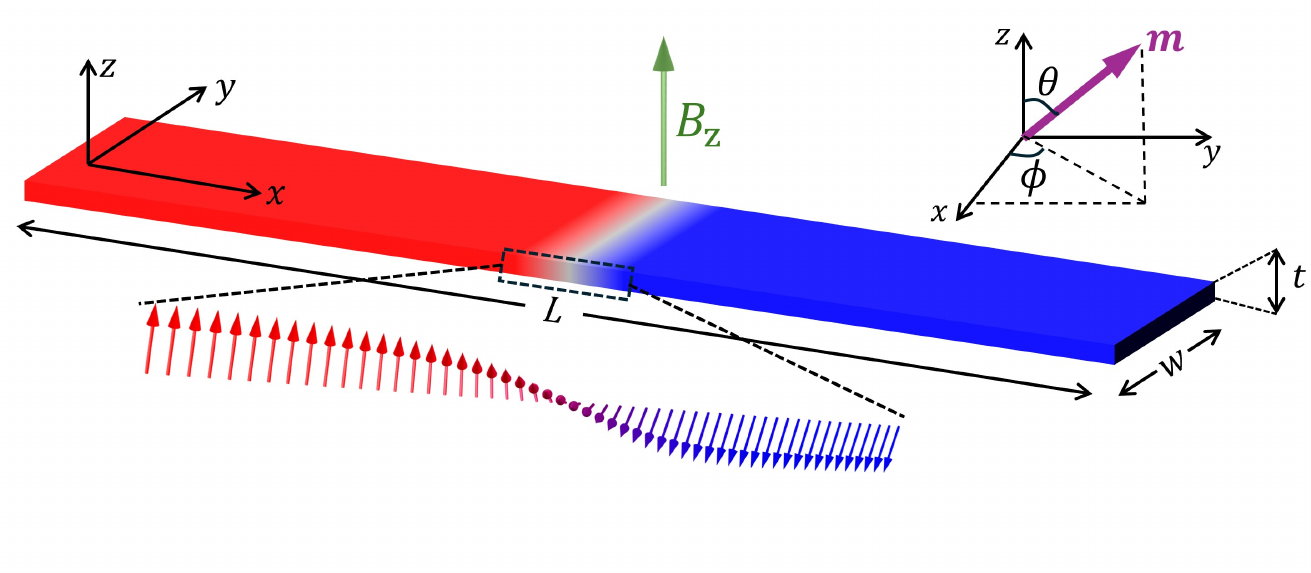}
    \caption{Schematic representation of the system under study. An ultrathin perpendicularly magnetized nanowire containing a domain wall at its center subjected to an out-of-plane external magnetic field $B_z$. The top-right inset defines the spherical coordinate system for the local magnetization $\bm{m}$.}
    \label{fig_geom}
\end{figure}

We consider a magnetic DW separating two opposing domains within a narrow nanowire composed of a perpendicularly magnetized magnetostrictive ferromagnet, as represented schematically in Fig.~\ref{fig_geom}. Our goal is to evaluate the dynamic response of this magnetic texture along the longitudinal axis of the nanowire under an externally applied magnetic field oriented along the $z$-direction, which drives the DW propagation to the right.

To capture the influence of mechanical degrees of freedom on DW dynamics, the system's time evolution requires the self-consistent, concurrent solution of the Landau-Lifshitz-Gilbert (LLG) equation (\ref{eq_LLG}) for the unitary magnetization vector $\bm{m}$ and the elastodynamic wave equation (\ref{eq_elastic_dynamics}) for the displacement field $\bm{u}$. The bidirectional coupling between these subsystems is fundamentally embedded in two terms: the magnetoelastic contribution to the effective magnetic field in Eq.~(\ref{eq_LLG}) and the magnetostrictive stress tensor contribution to the mechanical response in Eq.~(\ref{eq_elastic_dynamics}). The comprehensive three-dimensional coupled formalism is detailed in Appendix~\ref{sec_general_formalism}.

To circumvent the extensive computational overhead of the full 3D framework while preserving the core physical mechanisms, we reduce the general framework to a self-consistent, effective one-dimensional (1D) model. For a narrow, high-aspect-ratio nanowire extended along the $x$-axis, all state variables can be assumed to depend solely on the longitudinal coordinate $x$, causing spatial derivatives with respect to $y$ and $z$ to vanish identically ($\partial_y = \partial_z = 0$). To decouple transverse mechanical modes, we set the cross-coupling elastic stiffness constant $C_{12} = 0$ and impose that out-of-axis stress components vanish ($\sigma_{yy} = \sigma_{zz} = \sigma_{yz} = 0$). Physically, this implies that geometric strains instantaneously compensate for transverse magnetostriction ($\varepsilon_{ij}^\mathrm{el} = \varepsilon_{ij} - \varepsilon_{ij}^\mathrm{m} = 0$). While this formulation retains the transverse deformations induced by magnetostriction, their dynamic oscillations are structurally frozen by transverse clamping and rapid relaxation. Consequently, out-of-axis mechanical modes do not feed back into the magnetization dynamics, successfully isolating the primary longitudinal magnetoelastic interaction.

Under these geometric reductions, the magnetoelastic energy density [Eq.~(\ref{eq_me_energy})] simplifies to

\begin{equation}
    \mathcal{E}_{me} = \frac{1}{2}\sigma_{xx}\varepsilon_{xx}^\text{el} + \sigma_{xy}\varepsilon_{xy}^\text{el} +\sigma_{xz}\varepsilon_{xz}^\text{el} \, , 
\end{equation}

\noindent and the total energy density of the system reads

\begin{equation}
\label{eq_energy_density}
    \mathcal{E}_\text{tot} = A \left[ \left( \frac{\partial\theta}{\partial x}\right)^2 + \sin^2\theta \left(\frac{\partial\phi}{\partial x}\right)^2\right] + (K_0 + K_1 \sin^2\phi) \sin^2\theta - M_s B_z \cos\theta + \mathcal{E}_\mathrm{me} \, ,
\end{equation}

\noindent where $(\theta,\phi)$ denote the spherical polar and azimuthal angles of the magnetization vector (Fig.~\ref{fig_geom}), $M_s$ is the saturation magnetization, and $A$ is the exchange stiffness. The total effective anisotropy is partitioned into an out-of-plane contribution $K_0 = K_u + \frac{1}{2}\mu_0 M_s^2 (N_x-N_z)$ and an in-plane contribution $K_1 = \frac{1}{2}\mu_0 M_s^2 (N_y-N_x)$, $N_{x,y,z}$ being the diagonal demagnetizing factors of the nanowire cross-section. Here, $K_0$ combines the intrinsic perpendicular magnetocrystalline anisotropy $K_u$ with the out-of-plane magnetostatic demagnetizing energy, ensuring a perpendicular easy axis along $z$ ($K_0 > 0$), whereas $K_1$ represents a purely magnetostatic in-plane shape anisotropy that favors magnetization alignment along the wire axis ($K_1>0$).

The magnetoelastic contribution to the effective magnetic field [Eq.~(\ref{eq_me_field})] reduces to the vector field

\begin{equation}
\label{eq_B_me_1d}
    \bm{B}_\mathrm{me}  = 
    \frac{3}{M_s}  \left\{ \left[ \lambda_{100} \, \sigma_{xx} \, m_x + \lambda_{111} \, \left( \sigma_{xy} \, m_y + \sigma_{xz} \, m_z \right) \right] \, \hat{x} + \lambda_{111} \, m_x \, \left( \sigma_{xy} \hat{y} + \sigma_{xz} \hat{z} \right) \right\} \, ,
\end{equation}

\noindent where $\lambda_{100}$ and $\lambda_{111}$ are the saturation magnetostriction constants. 

The dynamic state of the material is governed by the active components of the stress tensor:

\begin{equation}
\label{eq_sigma}
\begin{split}
    \sigma_{xx} &= C_{11} \left[
    \frac{\partial u_x}{\partial x}-\frac{3}{2}\lambda_{100}\left( m_x^2 -\frac{1}{3} \right) \right]  \, , \\
    \sigma_{xy} &= C_{44} \left( \frac{\partial u_y}{\partial x}-3\lambda_{111}\, m_x m_y \right)  \, , \\
    \sigma_{xz} &= C_{44} \left( \frac{\partial u_z}{\partial x}-3\lambda_{111}\, m_x m_z \right) \, , \\
\end{split}
\end{equation}

\noindent where $C_{11}$ and $C_{44}$ are the elastic stiffness constants in Voigt notation. Substituting Eq.~(\ref{eq_sigma}) into the 1D elastodynamic equations (\ref{eq_elastic_dynamics}) yields the mechanical wave equation driving the displacement field $\bm{u}$:

\begin{equation}
\label{eq_elastic_dynamics_1d}
\begin{split}
    \rho\frac{\partial^2\bm{u}}{\partial t^2} + \rho\eta\frac{\partial\bm{u}}{\partial t}
    & = C_{11} \left(\frac{\partial^2u_x}{\partial x^2}-3\lambda_{100}\,m_x\frac{\partial m_x}{\partial x} \right) \hat{x} \\
    & + C_{44} \left(\frac{\partial^2u_y}{\partial x^2}-3\lambda_{111}\,\frac{\partial(m_x m_y)}{\partial x} \right) \hat{y} \\
    & + C_{44} \left(\frac{\partial^2u_z}{\partial x^2}-3\lambda_{111}\,\frac{\partial(m_x m_z)}{\partial x} \right) \hat{z}  \, ,
\end{split}
\end{equation}

\noindent where $\rho$ is the mass density and $\eta$ is the elastic damping rate.

To systematically investigate the strain-mediated DW dynamics, we perform 1D micromagnetic and elastodynamic numerical simulations on an ultrathin nanowire of length $L = 1\,\mu\text{m}$, width $w = 10\,\text{nm}$, and thickness $t = 1\,\text{nm}$ (Fig.~\ref{fig_geom}). The physical parameters used in our calculations are listed in Table~\ref{tab:material_parameters}. The intrinsic magnetic and elastic properties—including exchange stiffness $A$, saturation magnetization $M_s$, perpendicular anisotropy $K_u$, density $\rho$, and elastic moduli $C_{ij}$—are chosen to match experimental values for amorphous $\text{Co/Pt}$ thin films \cite{Fattouhi_25}.  To accentuate and isolate the role of elastodynamic coupling on DW motion, the magnetostriction constants ($\lambda_{100}, \lambda_{111}$) are artificially scaled up. Nevertheless, the resulting magnitude ($\lambda \sim 10^{-4}$) remains physically realistic and corresponds to values typical of highly magnetostrictive alloys such as $\text{FeGa}$ (Galfenol) or $\text{SmFe}_2$.

\begin{table}[htbp]
\centering
\caption{Dimensions and material parameter values used in the simulations.}
\label{tab:material_parameters}
\begin{tabular}{lccc}
\toprule
\textbf{Parameter} & \textbf{Symbol} & \textbf{Value} & \textbf{Units} \\
\midrule
Nanowire Length & $L$ & $1$ & $\mu\mathrm{m}$ \\
Nanowire Width & $w$ & $10$ & $\mathrm{nm}$ \\
Nanowire Thickness & $t$ & $1$ & $\mathrm{nm}$ \\
\midrule
Exchange Stiffness & $A$ & $30$ & $\mathrm{pJ/m}$ \\
Saturation Magnetization & $M_s$ & $0.58$ & $\mathrm{MA/m}$ \\
Uniaxial Anisotropy & $K_u$ & $0.9$ & $\mathrm{MJ/m^3}$ \\
Gilbert Damping Parameter & $\alpha$ & $0.06$ & --- \\
Mass Density & $\rho$ & $2.08 \times 10^4$ & $\mathrm{kg/m^3}$ \\
Elastic Stiffness Constant & $C_{11}$ & $298$ & $\mathrm{GPa}$ \\
Elastic Stiffness Constant & $C_{12}$ & $0$ & $\mathrm{GPa}$ \\
Elastic Stiffness Constant & $C_{44}$ & $133$ & $\mathrm{GPa}$ \\
Magnetostriction Constants & $\lambda_{100}, \lambda_{111}$ & $-5 \times 10^{-4}$ & --- \\
\bottomrule
\end{tabular}
\end{table}

To integrate the system numerically, the second-order mechanical wave equation [Eq.~(\ref{eq_elastic_dynamics_1d})] is recast as a set of two coupled first-order differential equations by introducing the mechanical velocity field $\bm{v} = \frac{\partial\bm{u}}{\partial t}$. The complete dynamical state of the nanowire is thus defined by nine coupled variables: three components for the magnetization $\bm{m}$, three for the mechanical displacement $\bm{u}$, and three for the velocity $\bm{v}$. Spatial derivatives are discretized using a second-order central finite-difference scheme. The resulting system of ordinary differential equations (ODEs) is integrated forward in time using an explicit second-order predictor-corrector Heun integration scheme with a constant temporal step $\Delta t$. At each time step, the magnetization vector field is strictly re-normalized ($\vert{}\bm{m}\vert{} = 1$) to enforce micromagnetic conservation constraints. Rigid mechanical clamping is applied at both boundaries of the wire ($x = \pm L/2$), setting the displacement field to zero:

\begin{equation}
\label{eq_boundary_conditions}
    \bm{u}\left(x = \pm L/2 \right) = 0.
\end{equation}

\subsection{Equilibrium solution}

Before investigating the dynamic response under an applied magnetic field, it is instructive to establish an analytical solution for the ground state of our coupled system in static equilibrium ($B_z = 0$). In mechanical equilibrium, the divergence of the stress tensor vanishes identically, $\nabla \cdot \bm{\sigma} = 0$ [Eq.~(\ref{eq_elastic_dynamics})], meaning that the active stress tensor components are uniform along the wire axis ($\partial_x \sigma_{xx} = \partial_x \sigma_{xy} = \partial_x \sigma_{xz} = 0$). Far from the DW core ($x \to \pm\infty$), the magnetization aligns uniformly along the out-of-plane easy axis ($m_z = \pm 1$, $m_x = m_y = 0$) with vanishing mechanical strain ($\partial_x u_i = 0$). Evaluating Eq.~(\ref{eq_sigma}) at these asymptotic limits yields: $\sigma_{xx} = \frac{1}{2} C_{11} \lambda_{100}$ and $\sigma_{xy} = \sigma_{xz} = 0$.

Substituting these stress values into Eq.~(\ref{eq_B_me_1d}) simplifies the static magnetoelastic effective field to $\bm{B}_\mathrm{me} = \frac{3 \lambda_{100}^2 C_{11}}{2 M_s} m_x \hat{x}$, which acts formally as an additional in-plane uniaxial anisotropy field oriented along the $x$-axis. Assuming a uniform internal azimuthal angle along the nanowire ($\phi(x) = \varphi$, $\partial_x \phi = 0$), the static total energy density [Eq.~(\ref{eq_energy_density})] reduces to

\begin{equation}
\label{eq_energy_density_equilibrium}
    \mathcal{E}_\text{tot} = A \left( \frac{d\theta}{dx}\right)^2 + \left[(K_0 - K_\mathrm{me}) + (K_1+K_\mathrm{me}) \sin^2\varphi \right] \sin^2\theta \, ,
\end{equation}

\noindent where $K_\mathrm{me} = \frac{3}{4} \lambda_{100}^2 C_{11}$ denotes the magnetoelastic anisotropy contribution. Minimizing the functional in Eq.~(\ref{eq_energy_density_equilibrium}) yields the familiar 1D DW profile $\theta(x) = 2 \tan^{-1} \left[ \exp \left( Q \frac{x}{\Delta} \right) \right]$, corresponding to

\begin{equation}
\label{eq_m_eq}
\begin{split}
    m_x^\text{eq}(x,\varphi) &= \text{sech}\left(\frac{x}{\Delta}\right) \, \cos\varphi \, , \\
    m_y^\text{eq}(x,\varphi) &= \text{sech}\left(\frac{x}{\Delta}\right) \, \sin\varphi \, , \\
    m_z^\text{eq}(x,\varphi) &= -Q\tanh\left(\frac{x}{\Delta}\right) \, ,
\end{split}
\end{equation}

\noindent in Cartesian components, where $Q = \pm 1$ is the topological wall charge and $\Delta(\varphi)$ represents the equilibrium DW width:

\begin{equation}
\label{eq_Delta_eq}
    \Delta(\varphi) = \left( \frac{A}{K_0 - K_\mathrm{me} + (K_1+K_\mathrm{me}) \sin^2\varphi} \right)^{1/2}.
\end{equation}

By substituting the magnetic ground-state profile [Eq.~(\ref{eq_m_eq})] into the static mechanical equations ($\partial_x \sigma_{xi} = 0$, $i = x,y,z$), we solve the second-order ordinary differential equations for each Cartesian component of the mechanical displacement field $\bm{u}$:

\begin{equation}
\label{eq_u_eq}
\begin{split}
    u_x^\text{eq}(x,\varphi) &= \frac{3}{2} \,\lambda_{100} \, \Delta \, \cos^2\varphi \,
    \left[\tanh\left(\frac{x}{\Delta}\right) - \frac{2x}{L} \right] \, , \\
    u_y^\text{eq}(x,\varphi) &= 3 \, \lambda_{111}\, \Delta \, \sin\varphi\,\cos\varphi\, \left[\tanh\left(\frac{x}{\Delta}\right) - \frac{2x}{L} \right] \, , \\
    u_z^\text{eq}(x,\varphi) &= 3 \, \lambda_{111}\,Q \, \Delta \, \cos\varphi \, \text{sech}\left(\frac{x}{\Delta}\right).
\end{split}
\end{equation}

The linear spatial terms ($-2x/L$) in $u_x$ and $u_y$ enforce the clamped boundary conditions [Eq.~(\ref{eq_boundary_conditions})] given that $\tanh(x/\Delta) \to \pm 1$ for $x \to \pm L \gg \Delta$. Differentiating Eq.~(\ref{eq_u_eq}) with respect to $x$ produces the corresponding equilibrium geometric strain components via Eq.~(\ref{eq_geometric_strain}):

\begin{equation}
\label{eq_eps_eq}
\begin{split}
    \varepsilon_{xx}^\text{eq}(x,\varphi) &= \frac{3}{2} \,\lambda_{100}  \, \cos^2\varphi \,
    \left[\text{sech}^2\left(\frac{x}{\Delta}\right) - \frac{2\Delta}{L} \right] \, , \\
    \varepsilon_{xy}^\text{eq}(x,\varphi) &= \frac{3}{2} \, \lambda_{111} \, \sin\varphi \, \cos\varphi\, \left[\text{sech}^2\left(\frac{x}{\Delta}\right) - \frac{2\Delta}{L} \right] \, , \\
    \varepsilon_{xz}^\text{eq}(x,\varphi) &= -\frac{3}{2} \, \lambda_{111} \,Q \, \cos\varphi \, \text{sech}\left(\frac{x}{\Delta}\right) \tanh\left(\frac{x}{\Delta}\right).
\end{split}
\end{equation}

\begin{figure} [!htb]
    \centering
    %\includesvg[width=1.0\linewidth]{fig_eq.svg}
    \includegraphics[width=1.0\linewidth]{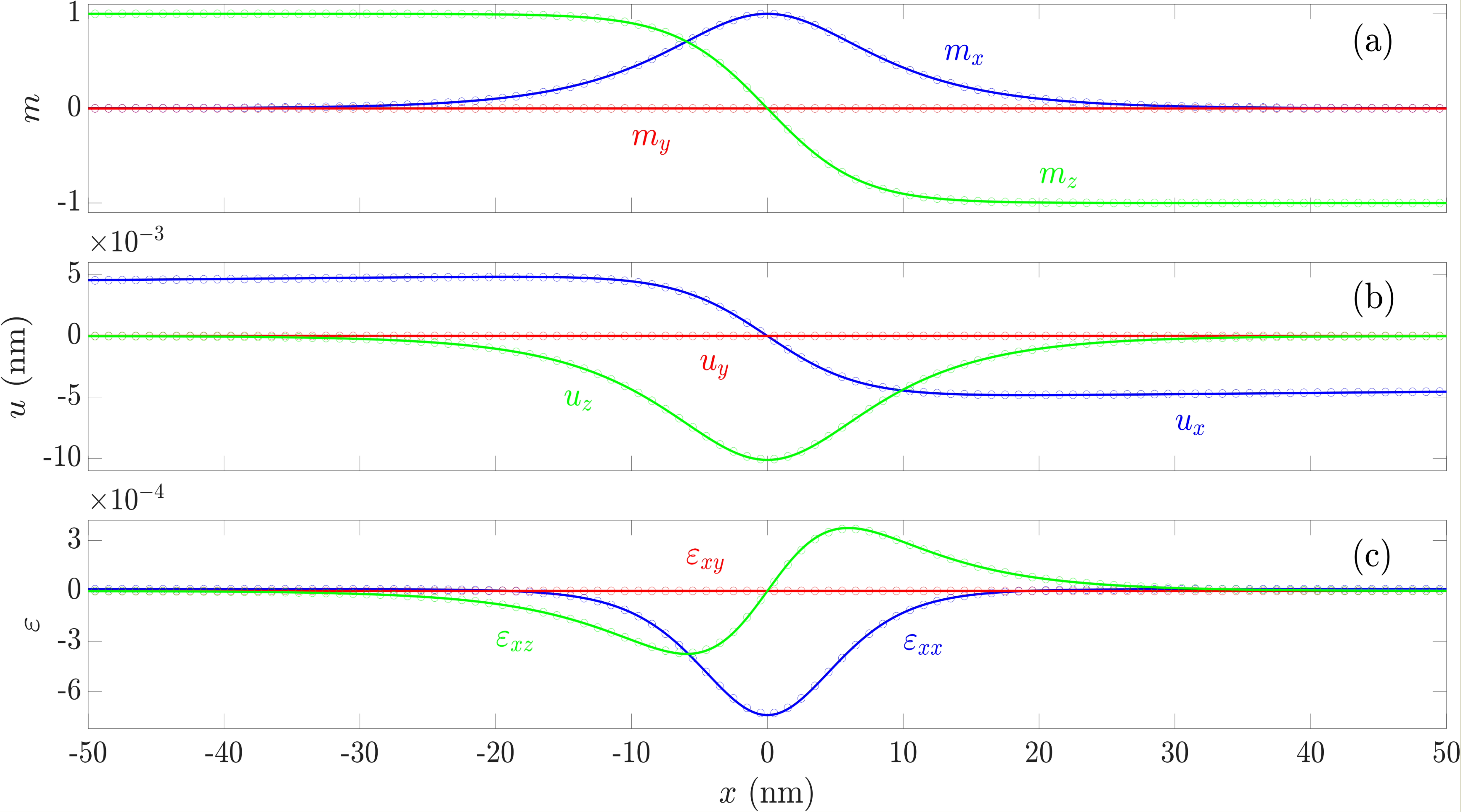}
    \caption{Equilibrium spatial profiles of magnetization (a), mechanical displacement (b), and geometric strain (c) in the absence of an applied magnetic field ($B_z = 0$). Solid lines represent analytical solutions [Eqs.~(\ref{eq_m_eq})--(\ref{eq_eps_eq})], and open circles represent numerical simulation results.}
    \label{fig_eq}    
\end{figure}

Because both the magnetostatic shape anisotropy $K_1$ and the magnetoelastic anisotropy $K_\mathrm{me}$ are strictly positive ($K_1, K_\mathrm{me} > 0$), energy minimization requires $\sin\varphi = 0$, selecting a ground-state Néel wall configuration ($\varphi = 0, \pi$). Figure~\ref{fig_eq} displays the resulting spatial profiles for magnetization (a), mechanical displacement (b), and geometric strain (c). The localized magnetic DW induces both longitudinal and transverse elastic deformations localized around the wall core over length scales comparable to $\Delta$. As shown in Fig.~\ref{fig_eq}, the analytical predictions (solid lines) agree well with micromagnetic-elastodynamic numerical simulations (open circles).

\section{Domain wall dynamics without elastic losses ($\eta = 0$)}
\label{sec_eta_0}

\subsection{Simulation results}

Having characterized the equilibrium DW configuration of the coupled magnetoelastic system, we turn our attention to the dynamic response driven by an out-of-plane external magnetic field ($B_z$). In this section, we focus exclusively on the regime devoid of mechanical dissipation by setting the elastic damping parameter to zero ($\eta = 0$), leaving its explicit effects to be detailed in Section~\ref{sec_eta_not_0}. To isolate and clearly quantify the consequences of the bidirectional magnetoelastic coupling, we benchmark the dynamic observables of our coupled system against a purely magnetic reference model where magnetoelastic interactions are omitted and only the standard spin dynamics are solved.

\begin{figure} [!htb]
    \centering
   % \includesvg[width=0.9\linewidth]{fig_e0_qvst.svg}
      \includegraphics[width=0.9\linewidth]{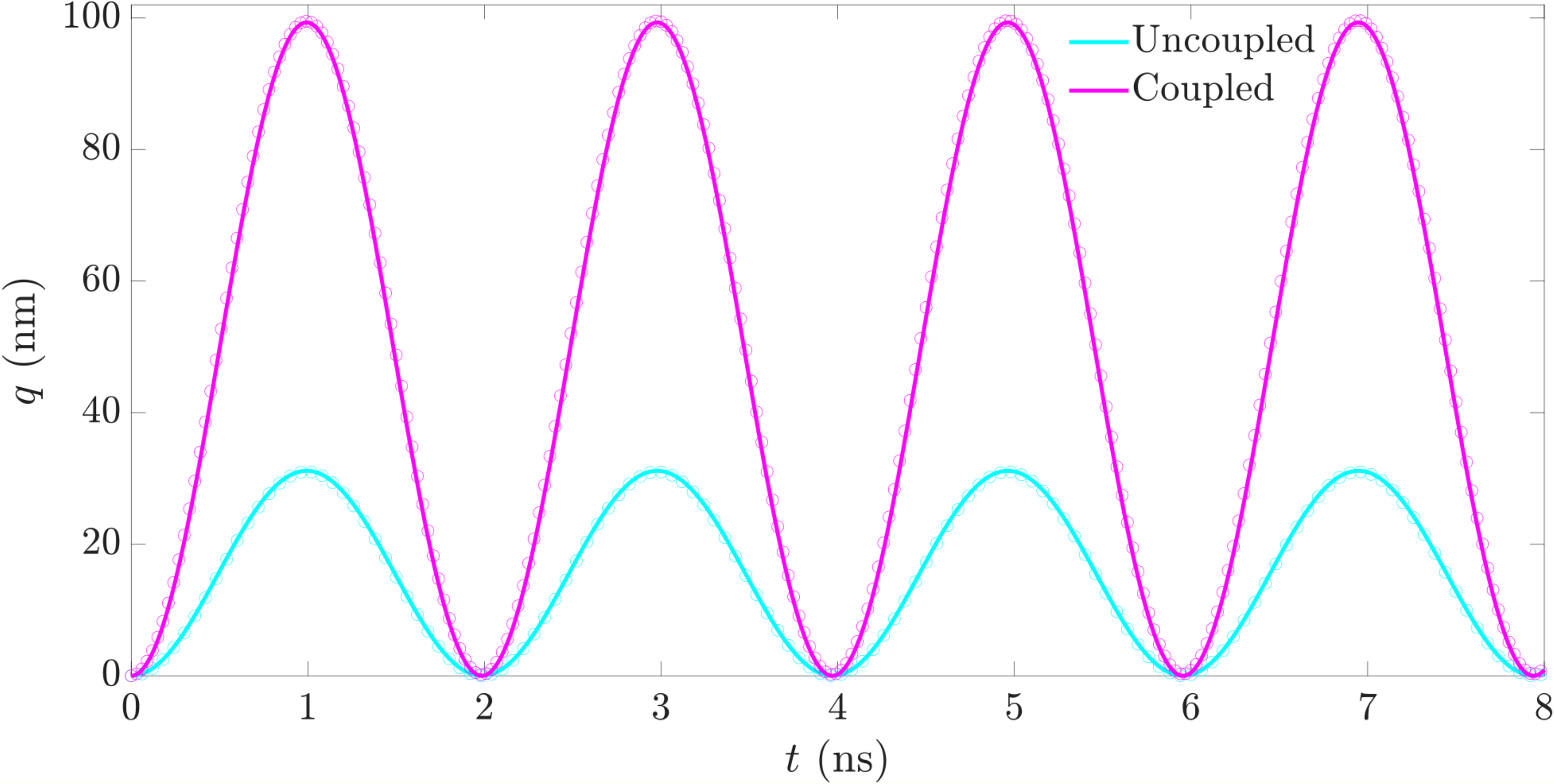}
    \caption{Time evolution of the DW position $q(t)$ under an applied magnetic field of $B_z = 9\,\text{mT}$ in the absence of both magnetic and mechanical dissipation ($\alpha = 0$, $\eta = 0$). The curves contrast the coupled magnetoelastic dynamics (magenta) with the uncoupled reference model (cyan). Solid lines represent the analytical solutions derived from Eqs.~(\ref{eq_Delta_eta_0}) and (\ref{eq_q_alpha_0}), while open circles denote the corresponding numerical simulation results.}
    \label{fig_e0_qvst}
\end{figure}

We begin our dynamic analysis by examining the ideal, dissipation-free regime where both mechanical and magnetic losses are entirely neglected ($\alpha = 0$, $\eta = 0$). Under these strict conservative conditions, the total energy of the system remains invariant, precluding any net, long-term translational motion of the DW. Figure~\ref{fig_e0_qvst} illustrates the transient DW position $q(t)$ as a function of time under an applied field $B_z = 9\,\text{mT}$. As anticipated, the DW exhibits sustained periodic oscillations with zero net displacement. Notably, while the characteristic oscillation frequency remains identical across both frameworks, the amplitude of the spatial oscillations is significantly enhanced in the coupled magnetoelastic system (magenta) compared to the uncoupled purely magnetic reference model (cyan). The physical origin of this amplitude boost will be clarified later through our collective coordinates framework.

\begin{figure}[!htb]
    \centering
    %\includesvg[width=1.0\linewidth]{fig_e0_vvsB.svg}
    \includegraphics[width=1.0\linewidth]{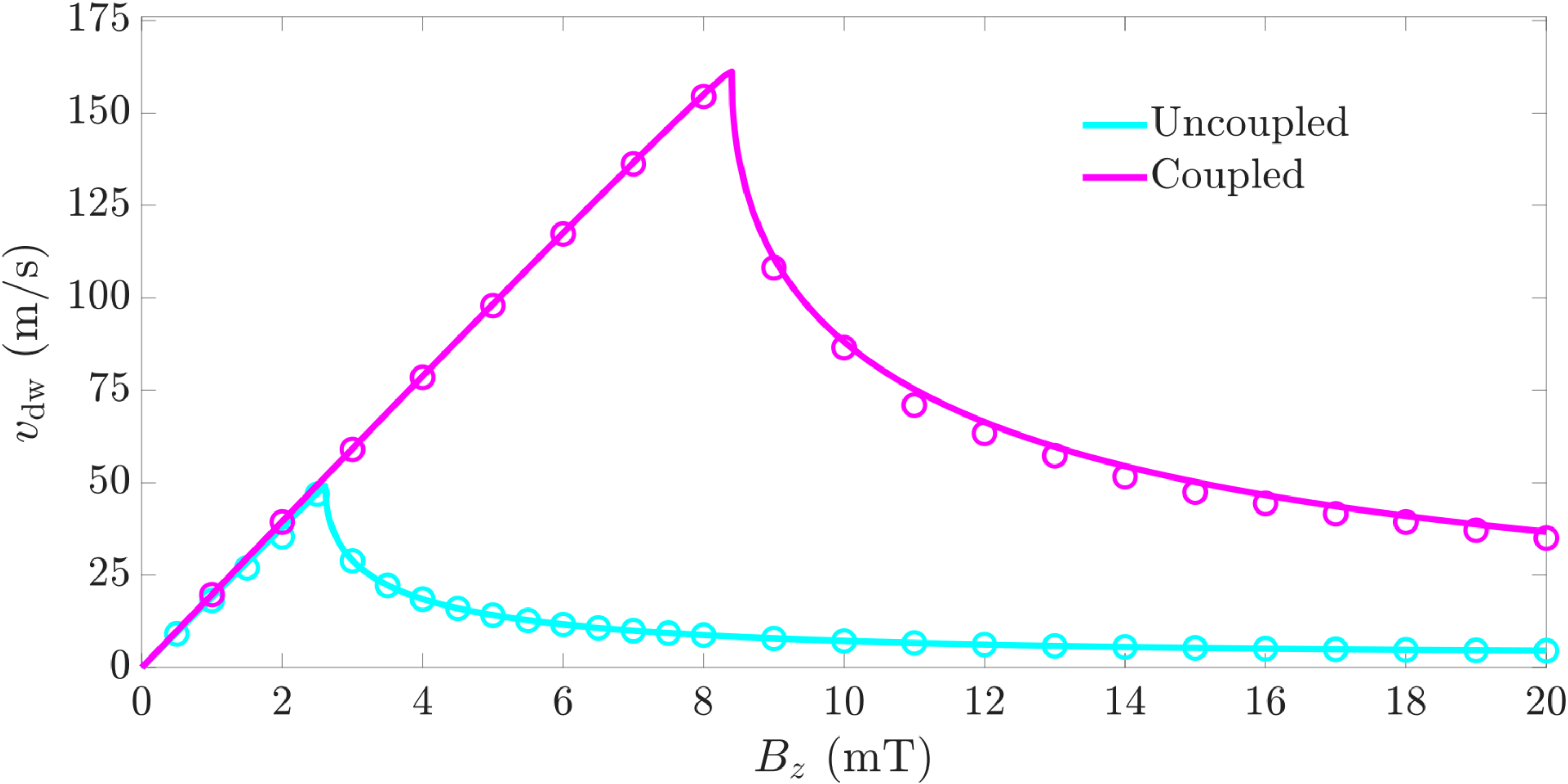}
    \caption{Average DW velocity as a function of the applied magnetic field for a finite Gilbert damping $\alpha = 0.06$ and zero mechanical dissipation ($\eta = 0$). Solid lines represent the theoretical predictions derived from Eqs.~(\ref{eq_q_dot_1}) and (\ref{eq_phi_dot_1}), while open circles denote the discrete numerical simulation results.}
    \label{fig_e0_vvsB}
\end{figure}

Next, we incorporate finite magnetic losses ($\alpha \neq 0$) while keeping the mechanical system dissipation-free ($\eta = 0$). In a purely magnetic system, the DW dynamics follow a well-established paradigm: under low driving fields, the wall undergoes rigid propagation with a steady velocity proportional to the external field and inversely proportional to $\alpha$. Above a critical threshold, the Walker breakdown field ($B_{\text{W}}$), rigid motion collapses, giving way to a precessional regime characterized by continuous in-plane magnetization rotation and a drastically reduced average forward velocity. Our numerical results for $\alpha = 0.06$ are compiled in Fig.~\ref{fig_e0_vvsB}, illustrating the average DW velocity as a function of the applied out-of-plane field. While the coupled magnetoelastic system qualitatively mirrors these two distinct dynamic phases, key quantitative deviations emerge due to the magnetoelastic interaction. Specifically, the coupled framework exhibits a slight enhancement in DW mobility within the linear regime, and the onset of the Walker breakdown is shifted toward a noticeably higher threshold field. 

It should be noted that within the precessional regime, the internal azimuthal rotation of the magnetization combined with the non-uniform DW motion induces the continuous emission of elastic waves that propagate away from the wall in both directions. Upon reaching the physical boundaries of the nanowire, these acoustic waves undergo reflection and travel backward, eventually interfering with the DW dynamics. To eliminate these spurious boundary reflections, we expanded the length of the simulation domain to $L = 6\,\mu\text{m}$ and carefully chose the temporal window to be sufficiently short to prevent reflected elastic wavefronts from re-entering the DW region, yet long enough to reliably compute the time-averaged velocity.

\subsection{Co-Moving Collective Coordinates Framework}

In order to understand the results for both $\alpha=0$ (Fig.~\ref{fig_e0_qvst}) and $\alpha\neq 0$ (Fig.~\ref{fig_e0_vvsB}), we consider the dynamic behavior of the elastic degrees of freedom. For $\eta = 0$, Eq.~(\ref{eq_elastic_dynamics_1d}) reduces to

\begin{equation}
    \label{eq_wave_equation}
    \frac{\partial^2 u_i}{\partial x^2} - \frac{1}{v_i^2}\frac{\partial^2 u_i}{\partial t^2} = p_i(x,t) \, ,
\end{equation}

\noindent an inhomogeneous wave equation for each Cartesian component of the displacement field $\bm{u}$, where $v_i$ represents the speed of sound in the medium (specifically, $v_x = v_\ell = \sqrt{C_{11}/\rho} = 3785\,\text{m/s}$ for longitudinal modes and $v_y=v_z=v_t=\sqrt{C_{44}/\rho} = 2529\,\text{m/s}$ for transverse modes). The source terms $p_i(x,t)$ are determined by spatial gradients of the magnetic DW profile [see Eq.~(\ref{eq_elastic_dynamics_1d})]. When the DW propagates rigidly at a constant velocity $v_{\text{dw}}$ far below the acoustic speed barrier ($v_{\text{dw}} \ll v_{\ell,t}$), the elastic displacement field adapts instantaneously to the traveling magnetic source~\cite{Achenbach_73}. Under this condition, the dynamic solution reduces to the equilibrium displacement field co-moving rigidly with the wall:

\begin{equation}
    \label{eq_u_adiabatic}
    \bm{u}(x,t) \approx \bm{u}^{\text{eq}}(x - v_{\text{dw}} t,\varphi).
\end{equation}

\begin{figure}[!htb]
    \centering
   % \includesvg[width=1.0\linewidth]{fig_e0_5mT.svg}
    \includegraphics[width=1.0\linewidth]{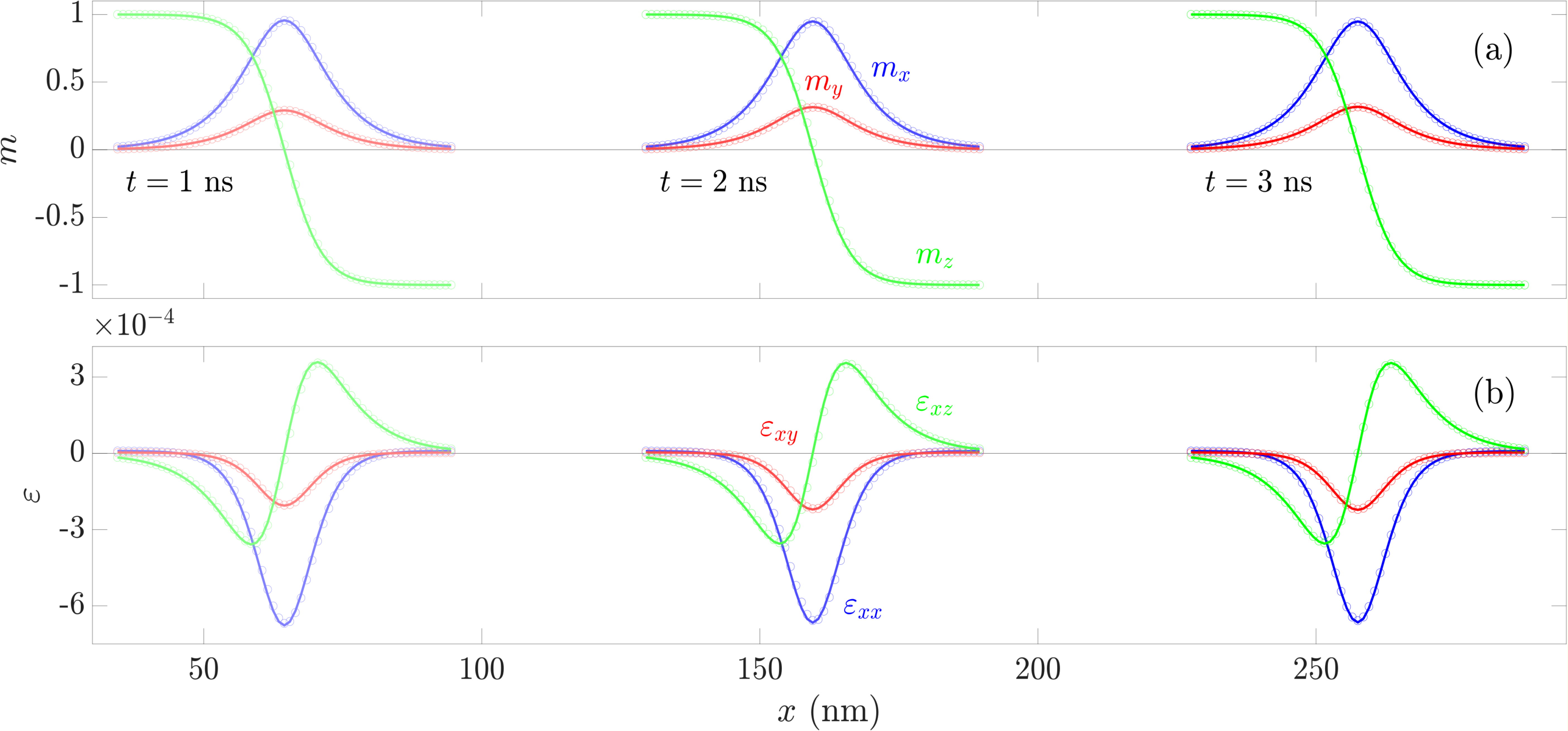}
    \caption{Spatial profiles of the (a) magnetization and (b) geometric strain components at three consecutive time instants ($t = 1$, $2$, and $3\,\text{ns}$) under an applied field of $B_z = 5\,\text{mT}$ ($\alpha = 0.06, \eta = 0$), corresponding to the steady-state regime (see Fig.~\ref{fig_e0_vvsB}). Successive time steps are depicted with progressively darker line intensities. Solid lines represent the analytical solutions within the adiabatic approximation, while open symbols denote the numerical simulation results. }
    \label{fig_e0_5mT}
\end{figure}

This regime is robustly fulfilled in our steady-state numerical simulations. As illustrated in Fig.~\ref{fig_e0_5mT}, both the magnetization components [Fig.~\ref{fig_e0_5mT}(a)] and the strain tensor profiles $\varepsilon_{xi}$ [Fig.~\ref{fig_e0_5mT}(b)] propagate without deformation under an applied field of $B_z = 5\,\text{mT}$ ($\alpha = 0.06$). We refer to this co-moving quasi-static state as the adiabatic regime, as the geometric strain slaves to the magnetic texture and retains its instantaneous equilibrium distribution.

\begin{figure}[!htb]
    \centering
   % \includesvg[width=1.0\linewidth]{fig_e0_9mT.svg}
     \includegraphics[width=1.0\linewidth]{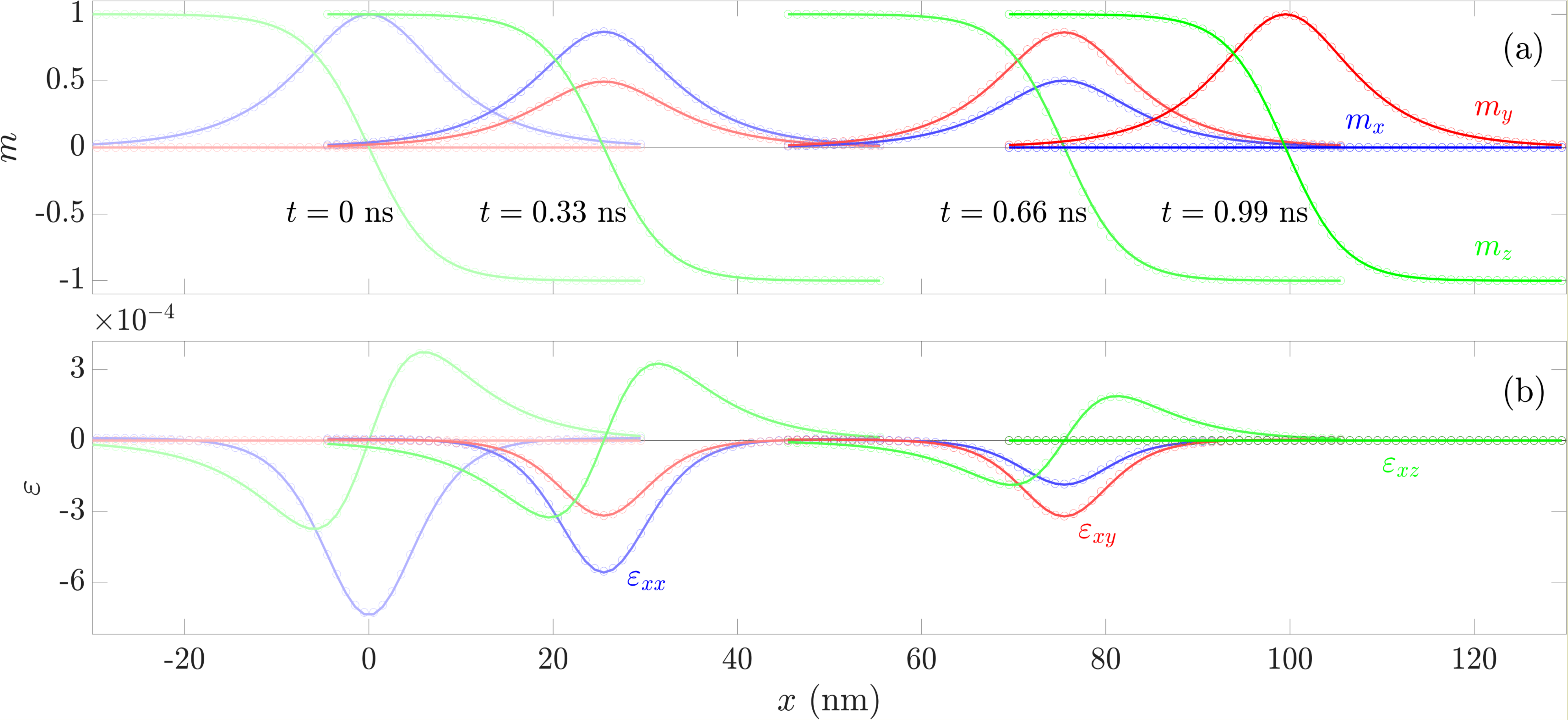}
    \caption{Spatial profiles of the (a) magnetization and (b) geometric strain components at four consecutive time instants ($t = 0,\, 0.33, \,0.66, \, 0.99 \, \text{ns}$) under an applied field of $B_z = 9\,\text{mT}$ ($\alpha = 0, \eta = 0$), corresponding to the precessional regime (see Fig.~\ref{fig_e0_qvst}). Successive time steps are depicted with progressively darker line intensities. Solid lines represent the analytical solutions derived within the adiabatic approximation, while open symbols denote the corresponding numerical simulation results.}
    \label{fig_e0_9mT}
\end{figure}

Remarkably, this adiabatic framework remains robust even within the precessional regime. Despite the continuous time evolution of the in-plane magnetization angle $\varphi$, and notwithstanding the aforementioned emission of elastic waves characteristic of this regime, the elastic deformation localized within the DW core dynamically follows the instantaneous structure of the moving wall. That is, the local strain continues to adhere to the adiabatic profile in Eq.~(\ref{eq_u_adiabatic}), governed at each time step by the instantaneous azimuthal angle $\varphi(t)$. This is validated in Fig.~\ref{fig_e0_9mT}, which displays the magnetization [Fig.~\ref{fig_e0_9mT}(a)] and strain profiles [Fig.~\ref{fig_e0_9mT}(b)] across four distinct phases of the precessional oscillation cycle in Fig.~\ref{fig_e0_qvst} ($B_z = 9\,\text{mT}, \alpha = 0$, $\eta=0$).

In view of these results, in the absence of mechanical dissipation ($\eta = 0$), we assume that the elastic displacement field co-moves with the DW according to $\bm{u}(x,t) = \bm{u}^\text{eq}[x - q(t), \varphi(t)]$, where $q(t)$ and $\varphi(t)$ denote the instantaneous DW position and internal angle, respectively. Differentiating this ansatz with respect to time generates inertial terms proportional to $\dot{q}^2$, $\ddot{q}$, $\dot{\varphi}^2$, and $\ddot{\varphi}$. However, because the DW velocity remains well below the speed of sound ($v_\text{dw} = \dot{q} \ll v_{\ell,t}$) and the elastic relaxation time across the DW core ($\tau_\text{el} \sim \Delta / v_{\ell,t} \sim \text{ps}$) is much shorter than the magnetic precessional period ($\tau_\text{mag} \sim \text{ns}$), the elastic displacement field responds adiabatically to changes in the magnetic state. Under these sub-sonic and adiabatic conditions, terms involving time derivatives of the DW parameters are strongly suppressed, simplifying the kinetic term in (\ref{eq_elastic_dynamics_1d}) to $\rho \frac{\partial^2 \bm{u}}{\partial t^2} \approx \rho \, v_\text{dw}^2 \frac{\partial^2 \bm{u}}{\partial x^2}$. Furthermore, since $\rho \, v_\text{dw}^2 \ll C_{ij}$, this inertial contribution is orders of magnitude smaller than the elastic restoring forces ($C_{ij} \frac{\partial^2 \bm{u}}{\partial x^2}$). Consequently, the elastodynamic equation reduces to the quasi-static mechanical balance $\nabla \cdot \bm{\sigma} \approx 0$ and therefore, as in static equilibrium, all normal and shear components of the total stress tensor remain spatially uniform ($\partial_x \sigma_{xx} = \partial_x \sigma_{xy} = \partial_x \sigma_{xz} = 0$). Applying the same boundary conditions as before yields $\sigma_{xx} = \frac{1}{2} C_{11} \lambda_{100}, \, \sigma_{xy} = \sigma_{xz} = 0$.

Thus, the effective magnetoelastic field [Eq.~(\ref{eq_B_me_1d})] again reduces to the simple form $\bm{B}_\mathrm{me}=\frac{3\lambda_{100}^2 C_{11}}{2 M_s} m_x\hat{x}$, which acts as an additional effective uniaxial anisotropy along the easy $x$-axis. Following the collective coordinates framework outlined in Appendix~\ref{sec_collective_coordinates}, we obtain the following coupled equations governing the time evolution of the DW position $q(t)$ and internal azimuthal angle $\varphi(t)$:

\begin{align}     
    \dot{q} &= \frac{\gamma\, Q \Delta \, }{1+\alpha^2} \left[ \frac{1}{2}(B_1+B_\sigma)\sin(2\varphi) + \alpha B_z \right] \label{eq_q_dot_1} \, , \\
    \dot{\varphi} &= \frac{\gamma}{1+\alpha^2} \left[ B_z-\frac{\alpha}{2} (B_1+B_\sigma)\sin(2\varphi) \right] \label{eq_phi_dot_1} \, ,
\end{align}

\noindent where $B_1=\frac{2K_1}{M_s}$ and $B_\sigma=\frac{2K_\text{me}}{M_s}$. Under this dynamic equilibrium, the instantaneous DW width adapts according to

\begin{equation}
\label{eq_Delta_eta_0}
    \Delta(\varphi) = \left( \frac{A}{K_0-K_\text{me} + (K_1+K_\text{me}) \sin^2\varphi(t)} \right)^{1/2}.
\end{equation}

\subsection{Discussion}

With these analytical expressions in hand, we can fully explain and accurately reproduce the physical behavior observed in Figs.~\ref{fig_e0_qvst} and \ref{fig_e0_vvsB}. In the undamped limit ($\alpha = 0$), the azimuthal velocity simplifies to $\dot{\varphi} = \gamma B_z$. This indicates a uniform internal precession at a constant frequency of $f = \frac{\gamma B_z}{2\pi} = 0.25\,\text{GHz}$, which remains identical for both the magnetoelastically coupled and uncoupled systems [Fig.~\ref{fig_e0_qvst}]. Substituting this uniform precessional rate into the velocity equation yields $\dot{q} = \gamma\, Q \, \Delta(\varphi) (B_1 + B_\sigma) \sin\varphi \cos\varphi$. Accounting for the instantaneous wall width ($\dot{\Delta}\approx 0$), the DW is given by:

\begin{equation}
    \label{eq_q_alpha_0}
    q(t) \approx \frac{Q\Delta[\varphi(t)]}{2} \frac{B_1+B_\sigma}{B_z} \sin^2\varphi(t).
\end{equation}

Equation~(\ref{eq_q_alpha_0}) shows that the DW oscillates periodically around its central position with an amplitude directly proportional to $(B_1 + B_\sigma)$ and inversely proportional to the applied field $B_z$. This analytical scaling clearly accounts for the enhancement of the amplitude in the coupled system compared to the uncoupled reference case in Fig.~\ref{fig_e0_qvst}, driven by the large magnetoelastic ratio ($B_\sigma / B_1 = 2.21$). As shown in Fig.~\ref{fig_e0_qvst}, the analytical trajectories match the numerical simulations with remarkable precision, a particularly striking agreement given that, as noted earlier, this precessional regime involves the continuous emission of elastic waves, which are entirely neglected within the collective coordinates framework.

We can similarly account for the velocity behavior shown in Fig.~\ref{fig_e0_vvsB}, where finite damping is included ($\alpha = 0.06$). In the steady-state regime, setting $\dot{\varphi} = 0$ in Eq.~(\ref{eq_phi_dot_1}) yields a constant internal equilibrium tilting angle $\varphi^*$ determined by

\begin{equation}
    \label{eq_varphi_eq}
    \sin(2\varphi^*) = \frac{2 B_z}{\alpha (B_1 + B_\sigma)} \, .
\end{equation}

\noindent Substituting Eq.~(\ref{eq_varphi_eq}) back into Eq.~(\ref{eq_q_dot_1}) results in a DW velocity that scales linearly with the applied magnetic field:

\begin{equation}
    \label{eq_v_linear}
    v_\text{dw} = \frac{\gamma \Delta(\varphi^*)}{\alpha} B_z \, .
\end{equation}

While the functional form of Eq.~(\ref{eq_v_linear}) is formally identical for both the coupled and uncoupled models, the coupled system exhibits a slightly higher mobility (slope $\mathrm{d}v_\text{dw}/\mathrm{d}B_z$) in Fig.~\ref{fig_e0_vvsB}. This enhancement stems directly from a larger equilibrium wall width $\Delta(\varphi^*)$. Indeed, according to Eq.~(\ref{eq_Delta_eta_0}), the DW expands in the coupled system due to two complementary effects: first, the explicit reduction of the effective anisotropy in the denominator via the $-K_{\text{me}}$ term; and second, the smaller equilibrium tilting angle $\varphi^*$ imposed by the additional magnetoelastic field in Eq.~(\ref{eq_varphi_eq}).

Furthermore, the Walker breakdown field $B_W$ is significantly elevated in the magnetoelastically coupled system [Fig.~\ref{fig_e0_vvsB}]. This shift follows directly from Eq.~(\ref{eq_varphi_eq}) by considering the limiting condition for steady-state motion, $\sin(2\varphi^*) = 1$, which gives

\begin{equation}
    \label{eq_B_W}
    B_W = \frac{\alpha}{2} (B_1 + B_\sigma).    
\end{equation}

Because the effective magnetoelastic field substantially reinforces the in-plane anisotropy ($B_\sigma / B_1 = 2.21$), the critical field required to initiate internal precession is increased by a factor of $(1 + B_\sigma/B_1) \approx 3.21$ relative to the uncoupled reference case, thereby extending the linear velocity regime to considerably higher driving fields.

In conclusion, the collective coordinates framework, formulated under the adiabatic co-moving approximation, provides excellent qualitative and quantitative accuracy for predicting DW dynamics in the zero-dissipation limit ($\eta = 0$). At first glance, this might suggest that simulations of the coupled system are redundant, as one could a priori assume spatially uniform stress components matching the boundary conditions and directly incorporate the resulting magnetoelastic energy into the magnetic 1D model. However, conducting the full coupled elastodynamic simulations was essential for two fundamental reasons. First, it rigorously validates that the adiabatic approximation correctly maps onto an effective magnetoelastic anisotropy field without losing crucial dynamic features. Second, and more importantly, resolving the full elastodynamic system reveals the physical mechanism underlying this behavior: the elastic continuum acts as a dynamic energy reservoir that actively exchanges energy with the magnetization.

\section{Domain wall dynamics with elastic losses ($\eta \neq 0$)}
\label{sec_eta_not_0}

\subsection{Simulation results}

We now turn to the core focus of this work: the impact of elastic damping on DW dynamics. Throughout this section, we set magnetic Gilbert damping to zero ($\alpha = 0$) unless explicitly noted otherwise.

\begin{figure}[!htb]
    \centering
    % \includesvg[width=1.0\linewidth]{fig_en0_qvst.svg}
     \includegraphics[width=1.0\linewidth]{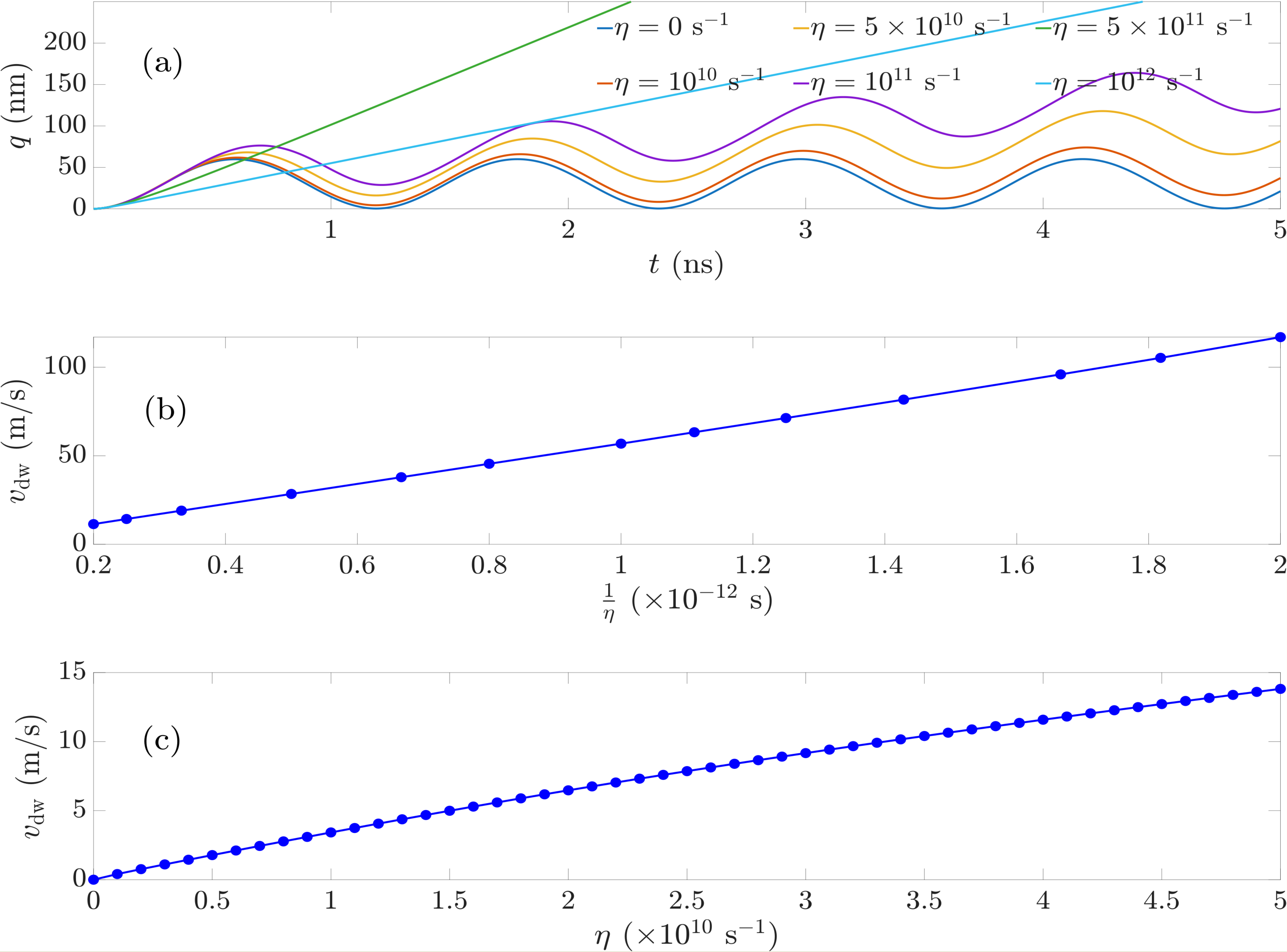}
    \caption{(a) Time evolution of the DW position $q(t)$ for $B_z = 15\,\text{mT}$ across different values of elastic damping $\eta$, obtained from numerical simulations. (b) Linear scaling of DW velocity with $\eta^{-1}$ in the steady regime ($\eta \ge 5\times 10^{11}\,\text{s}^{-1}$). (c) Non-linear dependence of the time-averaged DW velocity on $\eta$ in the precessional regime ($\eta < 5 \times 10^{11} \,\text{s}^{-1}$).}
    \label{fig_en0_qvst}
\end{figure}

As established in the previous section, when both magnetic and elastic dissipation are absent ($\alpha = \eta = 0$), total energy is strictly conserved. Under an applied magnetic field, the system undergoes periodic energy exchanges between the magnetic and elastic subsystems, forcing the DW to oscillate back and forth without achieving any net forward propagation [Fig.~\ref{fig_e0_qvst}]. The scenario changes drastically upon introducing finite mechanical damping ($\eta \neq 0$). This effect is illustrated in Fig.~\ref{fig_en0_qvst}(a), which shows the simulated time evolution of the DW position $q(t)$ under a constant drive $B_z = 15\,\text{mT}$ for various values of $\eta$. Clearly, the presence of elastic losses ($\eta \neq 0$) enables a net directional DW displacement. For low mechanical damping values, the wall enters a precessional regime. Here, it retains approximately the same precession frequency as in the undamped case, but acquires a constant net forward drift, represented by a finite mean slope in $q(t)$ that grows with increasing $\eta$ [Fig.~\ref{fig_en0_qvst}(c)]. As mechanical dissipation is increased further, reaching a threshold of $\eta \ge 5.0 \times 10^{11}\,\text{s}^{-1}$, precessional oscillations are completely suppressed. The wall instead settles into a steady propagation regime after a brief initial transient. In this steady regime, the velocity varies linearly with $\eta^{-1}$, as demonstrated in Fig.~\ref{fig_en0_qvst}(b). In contrast, within the precessional regime ($\eta < 5.0 \times 10^{11}\,\text{s}^{-1}$), the average velocity scales non-linearly with $\eta$ [Fig.~\ref{fig_en0_qvst}(c)].

These observations provide clear qualitative evidence that elastic dissipation acts as an effective drag mechanism for DW motion, playing a role remarkably analogous to intrinsic magnetic damping $\alpha$. This connection is naturally understood from a thermodynamic standpoint: mechanical damping opens a additional relaxation channel that continuously drains the Zeeman potential energy released by moving the DW forward.

\begin{figure}[!htb]
    \centering
    %\includesvg[width=1.0\linewidth]{fig_en0_vvsB.svg}
     \includegraphics[width=1.0\linewidth]{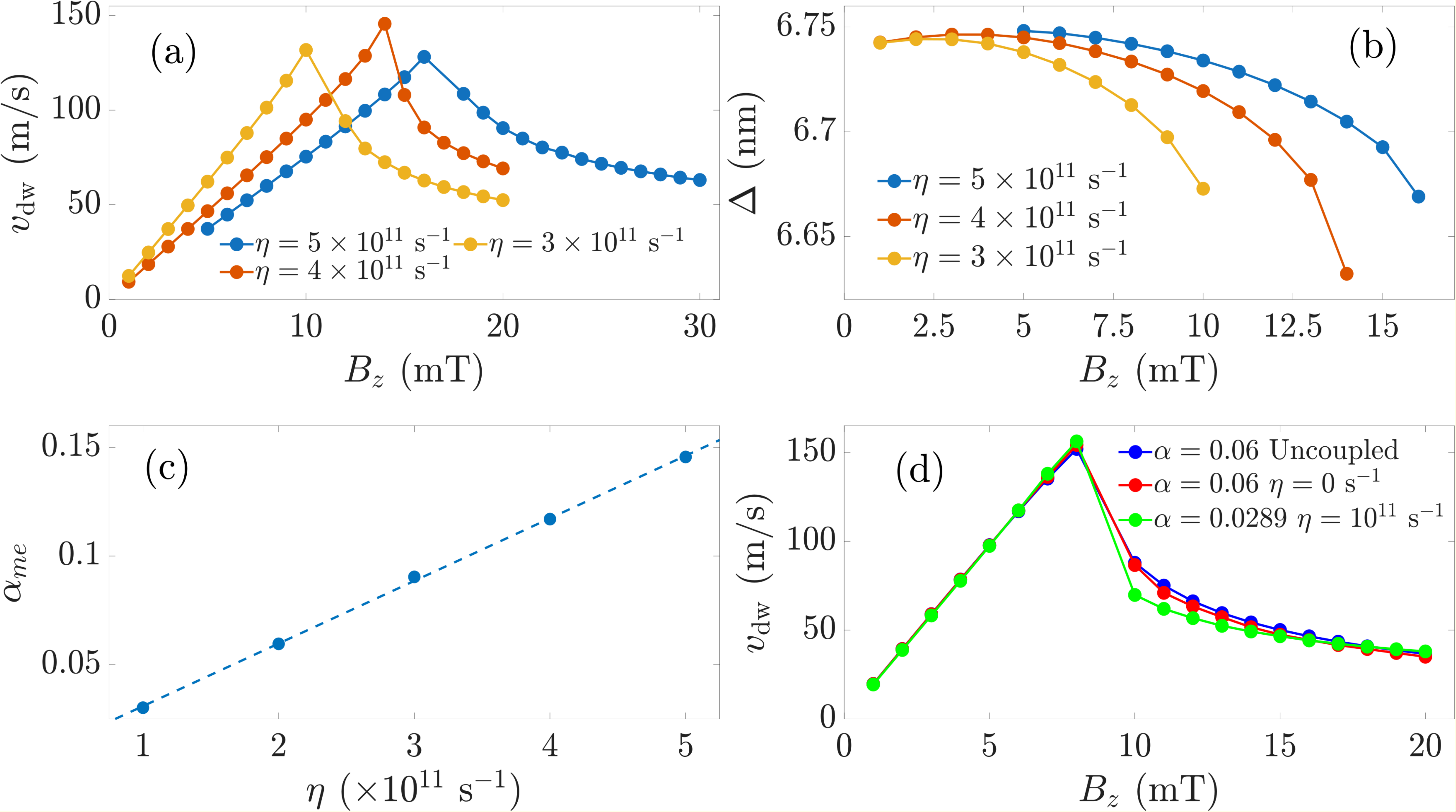}
    \caption{(a) DW velocity $v_\text{dw}$ and (b) DW width $\Delta$ as a function of the external magnetic field $B_z$ for different values of the elastic damping constant $\eta$ as obtained from numerical simulations. (c) Equivalent magnetoelastic damping constant $\alpha_\text{me}$ as a function of the elastic constant $\eta$ as computed by fitting the linear regime in (a) to $v=\gamma\Delta B_z / \alpha$. The dashed line corresponds to a linear fit to the data. (d) Comparison of DW velocity as a function of the applied field for three cases: uncoupled system with $\alpha=0.06$ and magnetoelastic contribution $\bm{B}_\mathrm{me}=\frac{3\lambda_{100}^2 C_{11}}{2 M_s} m_x\hat{x}$ (blue), coupled system with $\alpha=0.06$ and $\eta = 0$ (red) and coupled system with $\alpha=0.0289$ and $\eta=10^{11} \, \text{s}^{-1}$ (green), showing the equivalence of the three cases.}
    \label{fig_en0_vvsB}    
\end{figure}

Having established qualitative evidence that elastic damping $\eta$ affects DW motion in a manner analogous to magnetic damping $\alpha$, we now quantify this equivalence. Specifically, we determine the effective magnetic damping contribution $\alpha_{\text{me}}$ corresponding to a given value of $\eta$, and assess the precise extent to which these two loss mechanisms are interchangeable. To this end, we perform a systematic series of numerical simulations of DW dynamics as a function of the driving magnetic field $B_z$ for various values of $\eta$. Representative $v_\text{dw}(B_z)$ curves for three mechanical damping values are presented in Fig.~\ref{fig_en0_vvsB}(a). In all cases, the velocity response reproduces the classic features of damped micromagnetic dynamics: a steady regime at low fields characterized by a constant DW mobility $\mu = \mathrm{d}v_\text{dw}/\mathrm{d}B_z$, a distinct Walker breakdown field $B_W$ that shifts upward with increasing $\eta$, and a sharp drop in velocity for $B_z > B_W$. This striking similarity allows us to extract an equivalent magnetoelastic Gilbert damping constant $\alpha_{\text{me}}$. In the steady propagation regime, the DW velocity obeys $v_\text{dw} = \gamma \Delta B_z / \alpha$. By fitting the simulated low-field mobility slopes to this relation, we extract $\alpha_{\text{me}}$ for each value of $\eta$. In this fitting procedure, we account for the minor field dependence of the instantaneous DW width $\Delta(B_z)$ [Fig.~\ref{fig_en0_vvsB}(b)] by taking its mean value. This DW width is calculated numerically as $\Delta = \frac{1}{\pi} \int_{-L}^{+L} \sqrt{m_x^2 + m_y^2} \, \mathrm{d}x$.

Figure~\ref{fig_en0_vvsB}(c) displays the resulting effective damping $\alpha_{\text{me}}$ as a function of the mechanical damping parameter $\eta$. The relationship is strictly linear, yielding the simple constitutive relation:

\begin{equation}
    \label{eq_alpha_me}
    \alpha_\text{me} = 3.11 \times 10^{-13} \,\eta\thinspace(\text{s}^{-1}) .
\end{equation}

Equation~(\ref{eq_alpha_me}) represents one of the central findings of this work, as it directly maps mechanical energy dissipation into an effective magnetic Gilbert damping constant. To test the validity of this relation, Fig.~\ref{fig_en0_vvsB}(d) compares $v_\text{dw}(B_z)$ curves across three distinct simulation configurations: (i) an uncoupled system with total magnetic damping $\alpha = 0.060$ and an effective magnetoelastic anisotropy field $\bm{B}_{\text{me}} = \frac{3\lambda_{100}^2 C_{11}}{2 M_s} m_x \hat{x}$ (blue curve), (ii) a coupled system with purely magnetic dissipation $\alpha = 0.060$ and $\eta = 0$ (red curve) and (iii) a coupled system with mixed dissipation, combining intrinsic magnetic damping $\alpha = 0.0289$ and mechanical damping $\eta = 10^{11}\,\text{s}^{-1}$ (green curve), which according to Eq.~(\ref{eq_alpha_me}), contributes $\alpha_{\text{me}} = 0.0311$ (such that $\alpha_{\text{total}} = 0.0289 + 0.0311 = 0.0600$). As demonstrated in Fig.~\ref{fig_en0_vvsB}(d), the velocity-field characteristics of all three systems are virtually identical. While the equivalence between the first two cases was demonstrated in Section~\ref{sec_eta_0}, the close agreement with the third case strongly supports the validity of Eq.~(\ref{eq_alpha_me}). Furthermore, it demonstrates that magnetoelastic mechanical losses act as an additive dissipation channel that operates independently alongside standard phenomenological Gilbert damping.

\subsection{Co-Moving Collective Coordinates Framework}

To elucidate the physical mechanism driving the DW forward under mechanical dissipation, in what follows we develop an analytical framework based on an adiabatic co-moving approximation for the elastic continuum in the steady-state regime. 

\begin{figure}[!htb]
    \centering
   % \includesvg[width=1.0\linewidth]{fig_en0_sigma.svg}
     \includegraphics[width=1.0\linewidth]{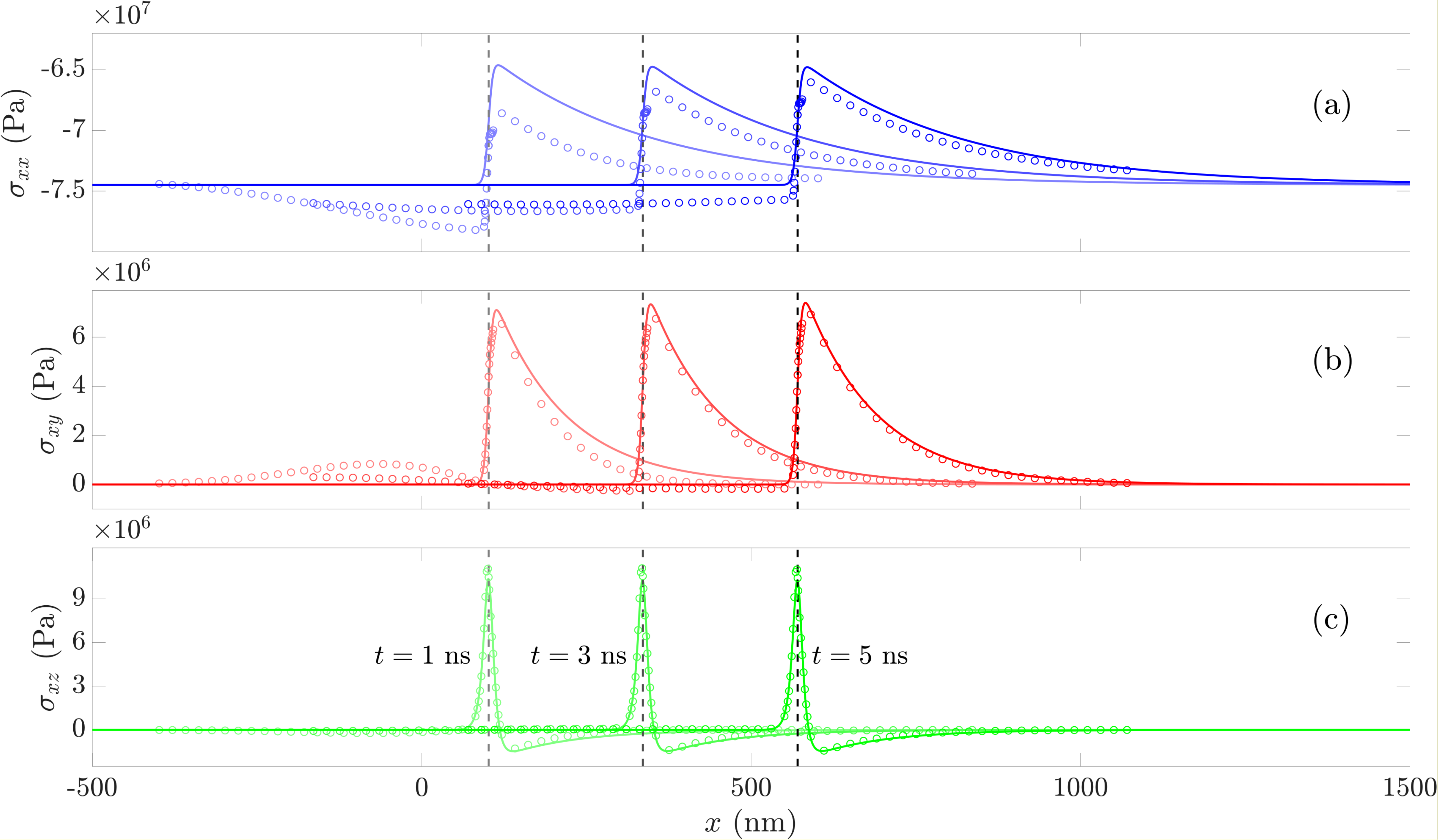}
    \caption{Spatial profiles of the components of the stress tensor (a) $\sigma_{xx}$ (b) $\sigma_{xy}$ and ($\sigma_{xz}$) at three consecutive time instants ($t=1, 3, 5 \, \text{ns}$) under an applied field of $B_z = 15 \, \text{mT}$ and $\eta = 5 \times 10^{11} \, \text{s}^{-1}$ ($\alpha = 0$), corresponding to the steady-state regime [see Fig.~\ref{fig_en0_qvst}(a)]. Successive time steps are depicted with progressively darker line intensities. Solid lines represent the analytical solutions within the adiabatic approximation, while open symbols denote the numerical simulation results. The position of the DW center at each time step is marked with a vertical dashed line.}
    \label{fig_en0_sigma}
\end{figure}

We first investigate the resulting spatial stress field in the elastic continuum. Figure~\ref{fig_en0_sigma} displays the spatial profiles of the stress tensor components at three consecutive time steps during steady-state propagation ($B_z = 15\,\text{mT}$, $\eta = 5 \times 10^{11}\,\text{s}^{-1}$, $\alpha = 0$). Unlike the lossless case ($\eta = 0$), where the stress tensor components remain strictly uniform in space, mechanical dissipation breaks this spatial symmetry. As a result, the stress tensor components develop characteristic non-uniform profiles around their respective equilibrium values ($\sigma_{xx}^\text{eq}=\frac{1}{2}C_{11}\lambda_{100}=-7.45 \times 10^7 \, \text{Pa}$, $\sigma_{xy}^\text{eq}=\sigma_{xz}^\text{eq}=0$). Specifically, $\sigma_{xx}$ and $\sigma_{xy}$ exhibit a positive gradient centered at the domain wall, relaxing gradually downstream along the forward propagation direction, whereas $\sigma_{xz}$ displays a pronounced positive peak localized at the domain wall core flanked by an asymmetric negative tail.

Building on these observations, in what follows we develop a reduced analytical framework based on an adiabatic co-moving approximation for the elastic fields. This model will allow us to quantify how these localized stress profiles generate the effective magnetoelastic forces that propel the DW forward.

In the presence of mechanical dissipation, the 1D elastodynamic wave equations for the displacement field components $u_i(x,t)$ [Eq.~(\ref{eq_elastic_dynamics_1d})] can be written as

\begin{equation}
    \label{eq_wave_equation_eta_not_0}
    \frac{\partial^2 u_i}{\partial x^2} - \frac{1}{v_i^2} \frac{\partial^2 u_i}{\partial t^2} - \frac{\eta}{v_i^2} \frac{\partial u_i}{\partial t}  = p_i(x,t) \, , 
\end{equation}

\noindent where, as mentioned in Eq.~(\ref{eq_wave_equation}), $v_i$ denotes the sound speed for the corresponding acoustic mode ($v_x = v_\ell$, $v_y = v_z = v_t$), and $p_i(x,t)$ represents the magnetoelastic source term determined by spatial gradients of the magnetization texture. As evidenced by Fig.~\ref{fig_en0_sigma}, the elastic deformation moves quasi-rigidly alongside the DW. We therefore invoke the co-moving traveling wave transformation $\xi = x - v_{\text{dw}}t$, setting $u_i(x,t) = u_i(\xi)$. Transforming derivatives via $\partial_x = \partial_\xi$ and $\partial_t = -v_{\text{dw}}\partial_\xi$ converts Eq.~(\ref{eq_wave_equation_eta_not_0}) into the ordinary differential equation

\begin{equation}
    \left[ 1 - \left( \frac{v_{\text{dw}}}{v_i} \right)^2 \right] \frac{\mathrm{d}^2 u_i}{\mathrm{d}\xi^2} + \frac{\eta v_{\text{dw}}}{v_i^2} \frac{\mathrm{d}u_i}{\mathrm{d}\xi} = p_i(\xi).
\end{equation}

\noindent In the sub-acoustic regime ($v_{\text{dw}} \ll v_i$), we neglect $(v_{\text{dw}}/v_i)^2 \ll 1$. Recognizing that the static equilibrium profile satisfies $\frac{\mathrm{d}^2 u_i^{\text{eq}}}{\mathrm{d}\xi^2} = p_i(\xi)$, the governing equation simplifies to

\begin{equation}
    \label{eq_u_i_ode}
    \frac{\mathrm{d}^2 u_i}{\mathrm{d}\xi^2} + \beta_i \frac{\mathrm{d}u_i}{\mathrm{d}\xi} = \frac{\mathrm{d}^2 u_i^{\text{eq}}}{\mathrm{d}\xi^2},
\end{equation}

\noindent where $\beta_i \equiv \frac{\eta v_{\text{dw}}}{v_i^2}$ characterizes the spatial decay length scale associated with viscous mechanical losses. Integrating Eq.~(\ref{eq_u_i_ode}) under vanishing displacement and strain conditions at infinity [Eq.~(\ref{eq_boundary_conditions})] yields 

\begin{equation}
    \label{eq_du_i_adiabatic_eta_not_0}
    \frac{\mathrm{d}u_i(\xi)}{\mathrm{d}\xi} + \beta_i u_i(\xi) = \frac{\mathrm{d}u_i^{\text{eq}}(\xi)}{\mathrm{d}\xi}.    
\end{equation}

\noindent When multiplying by the integrating factor $\mathrm{e}^{\beta_i \xi}$ and integrating by parts, the formal closed-form solution for the co-moving displacement field:

\begin{equation}
    \label{eq_u_i_adiabatic_eta_not_0}
    u_i(\xi) = u_i^{\text{eq}}(\xi) - \beta_i \mathrm{e}^{-\beta_i \xi} \int_{-\infty}^\xi u_i^{\text{eq}}(y) \mathrm{e}^{\beta_i y} \mathrm{d}y,    
\end{equation}

\noindent is obtained. Substituting Eq.~(\ref{eq_du_i_adiabatic_eta_not_0}) into Eq.~(\ref{eq_sigma}) directly provides the spatially resolved stress tensor components:

\begin{equation}
    \label{eq_sigma_xi}
    \sigma_{xi}(\xi) = \sigma_{xi}^{\text{eq}} - C_i \beta_i u_i(\xi),
\end{equation}

\noindent where $C_x = C_{11}$, $C_y = C_z = C_{44}$, and $u_i(\xi)$ is given by Eq.~(\ref{eq_u_i_adiabatic_eta_not_0}).

Equations~(\ref{eq_u_i_adiabatic_eta_not_0}) and (\ref{eq_sigma_xi}) provide explicit analytical stress distributions derived entirely from the known static equilibrium profiles $u_i^{\text{eq}}(\xi)$, although the integral in Eq.~(\ref{eq_u_i_adiabatic_eta_not_0}) has to be evaluated numerically. Furthermore, because the parameter product $\eta v_{\text{dw}}$ remains constant in the steady propagation regime [see Fig.~\ref{fig_en0_qvst}(b)], $\beta_i$ is uniquely determined without requiring prior knowledge of $v_{\text{dw}}$. As demonstrated in Fig.~\ref{fig_en0_sigma}, the analytical stress profiles (solid curves) show excellent agreement with elastodynamic simulations (open symbols). This confirms that our dissipative adiabatic model accurately captures the spatial distortion of the stress induced by mechanical damping. Minor discrepancies, mostly originated due to the finite-length domain in the simulations, decay progressively as the DW propagates.

To elucidate how these non-uniform stress profiles generate effective forces that drive the DW forward, we formulate a 1D collective coordinates model in the presence of mechanical losses.  From Eq.~(\ref{eq_B_me_1d}), the local magnetoelastic field spherical components are given by

\begin{subequations}
\label{eq_B_me_theta_phi}
\begin{align}    
    B_\text{me}^\theta &= \frac{3}{M_s} \left[ \left( \lambda_{100} \sigma_{xx} \cos^2\varphi + \lambda_{111} \sigma_{xy} \sin{2\varphi} \right) \sin\theta\cos\theta + \lambda_{111}\sigma_{xz}\cos\varphi \cos{2\theta} \right] \label{eq_B_me_theta} \, , \\
    B_\text{me}^\phi &= \frac{3}{M_s} \left[ \left( -\lambda_{100} \sigma_{xx} \sin\varphi\cos\varphi + \lambda_{111} \sigma_{xy} \cos{2\varphi} \right) \sin\theta - \lambda_{111}\sigma_{xz}\sin\varphi \cos\theta \right]  \, ,\label{eq_B_me_phi} 
\end{align}    
\end{subequations}

\noindent where the stress components $\sigma_{xi}$ obey Eqs.~(\ref{eq_u_i_adiabatic_eta_not_0}) and (\ref{eq_sigma_xi}). To integrate these fields into the collective coordinates framework, we invoke the standard 1D Walker profile ansatz [Eq.~(\ref{eq_theta_ansatz})], defining the local wall width parameter as

\begin{equation}
    \Delta = \sqrt{\frac{A}{K_\text{eff}}}.
\end{equation}

\noindent Notice that the stress components $\sigma_{xx}$ and $\sigma_{xy}$ in $B_\text{me}^\theta$ share the exact functional dependence ($\sin\theta\cos\theta$) of the intrinsic in-plane anisotropy $K_1$. Consequently, their contribution can be directly absorbed into a spatially dependent effective anisotropy $K_{\text{eff}}(x)$:

\begin{equation}
    K_\text{eff} = K_0 + K_1 \sin^2\varphi - \frac{3}{2}\lambda_{100} \,\sigma_{xx}\cos^2\varphi - \frac{3}{2}\lambda_{111} \,\sigma_{xy}\sin (2\varphi) .
\end{equation}

Conversely, the $\sigma_{xz}$ term exhibits a $\cos(2\theta)$ angular dependence, which cannot be incorporated into a standard scalar $K_\text{eff}$ without breaking the rigid Walker profile structure. Within our collective coordinates approach, we treat the Walker profile as a rigid baseline and omit the $\sigma_{xz}$-induced profile distortions in the definition of $\Delta$. As will be shown below, this leading-order approximation preserves analytical tractability while capturing the dominant dynamics observed in the numerical simulations with high fidelity.

Because the stress tensor components vary spatially, evaluating the exchange field requires taking into account the spatial gradient of the wall width, $\partial_x \Delta$. The first spatial derivative of $\Delta(x)$ is evaluated as

\begin{equation}
    \frac{\partial\Delta}{\partial x} = \frac{3 \,\Delta}{4}\frac{\lambda_{100} \,\sigma_{xx}'\cos^2\varphi + \lambda_{111} \,\sigma_{xy}' \sin 2\varphi}{K_\text{eff}}  \, ,
\end{equation}

\noindent where primes denote spatial derivatives ($\sigma_{ij}' \equiv \partial_x \sigma_{ij}$). Considering that spatial variations of $\Delta(x)$ occur over length scales much larger than the DW width, i.e., $\partial_x \Delta\cdot(x-q)\ll\Delta$ for $x\sim q$, we can approximate $\frac{\partial\theta}{\partial x} = \frac{Q}{\Delta}\sin\theta$, and the corresponding second derivative of the polar angle profile $\theta(x)$ simplifies to

\begin{equation}
    \label{eq_d2theta_dx2}
    \frac{\partial^2\theta}{\partial x^2} = \frac{1}{\Delta^2}\sin\theta\cos\theta - Q\frac{3}{4\Delta}\frac{\lambda_{100} \,\sigma_{xx}'\cos^2\varphi + \lambda_{111} \,\sigma_{xy}' \sin 2\varphi}{K_\text{eff}}.
\end{equation}

Inserting Eq.~(\ref{eq_d2theta_dx2}) alongside Eqs.~(\ref{eq_B_me_theta_phi}) and Eqs.~(\ref{eq_eq_B_eff_theta_and_phi_2}) into the torque equations [Eqs.~(\ref{eq_theta_and_phi_dot})] for zero magnetic damping ($\alpha = 0$) yields the local spatial evolution equations for $\dot{\theta}$ and $\dot{\varphi}$:

\begin{align}     
    \dot{\theta} &= - \frac{\gamma}{M_s}  \left[ \left( K_1 \sin 2\varphi + 3\lambda_{100} \, \sigma_{xx}\sin\varphi \cos\varphi - 3\lambda_{111} \, \sigma_{xy}\cos 2\varphi \right) \sin\theta + 3\lambda_{111} \sigma_{xz} \sin\varphi\cos\theta \right] \label{eq_theta_dot_eta_not_0} \, , \\
    \sin\theta \, \dot{\varphi} &=  \gamma \left[\ B_z \sin\theta + \tfrac{3 Q \Delta}{2 M_s} \left(\lambda_{100} \,\sigma_{xx}' \cos^2\varphi + \lambda_{111} \,\sigma_{xy}' \sin 2\varphi \right) \sin\theta - \tfrac{3}{M_s} \lambda_{111} \, \sigma_{xz} \cos \varphi \cos2\theta \right]. \label{eq_varphi_dot_eta_not_0}
\end{align}

In the steady-state propagation regime, the DW moves rigidly, meaning that the internal azimuthal angle reaches a stationary value $\varphi^*$ ($\dot{\varphi} = 0$). Imposing this condition in Eq.~(\ref{eq_varphi_dot_eta_not_0}) leads to an explicit  relation connecting the external driving field $B_z$ to the stress tensor components and their gradients:

\begin{equation}
    \label{eq_f}
    B_z \sin\theta = -\frac{3 Q \Delta}{2 M_s} \left( \lambda_{100} \sigma_{xx}' \cos^2\varphi^* + \lambda_{111} \sigma_{xy}' \sin 2\varphi^* \right)\sin\theta + \frac{3}{M_s} \lambda_{111} \sigma_{xz} \cos\varphi^* \cos 2\theta.
\end{equation}

\noindent Projecting this relation onto the DW structure by multiplying each term by $\sin\theta$ and integrating over $x$, we derive the integrated steady-state balance equation:

\begin{equation}
    \label{eq_f2}
    B_z = -\frac{3 Q \Delta}{2 M_s} \left( \lambda_{100} \langle \sigma_{xx}'\rangle \cos^2\varphi^* + \lambda_{111} \langle\sigma_{xy}'\rangle \sin 2\varphi^* \right) + \frac{3}{M_s} \lambda_{111} \langle\sigma_{xz}\rangle \cos\varphi^*,
\end{equation}

\noindent where the wall-averaged stress quantities are defined as:

\begin{equation}
    \langle \sigma_{xx}' \rangle = \frac{\int \sigma_{xx}' \sin^2\theta \, \mathrm{d}x}{\int \sin^2\theta \, \mathrm{d}x}, \quad \langle \sigma_{xy}' \rangle = \frac{\int \sigma_{xy}' \sin^2\theta \, \mathrm{d}x}{\int \sin^2\theta \, \mathrm{d}x}, \quad \langle \sigma_{xz} \rangle = \frac{\int \sigma_{xz} \cos(2\theta) \sin\theta \, \mathrm{d}x}{\int \sin^2\theta \, \mathrm{d}x}.
\end{equation}

\begin{figure}[!htb]
    \centering
   % \includesvg[width=0.9\linewidth]{fig_en0_fg.svg}
     \includegraphics[width=0.9\linewidth]{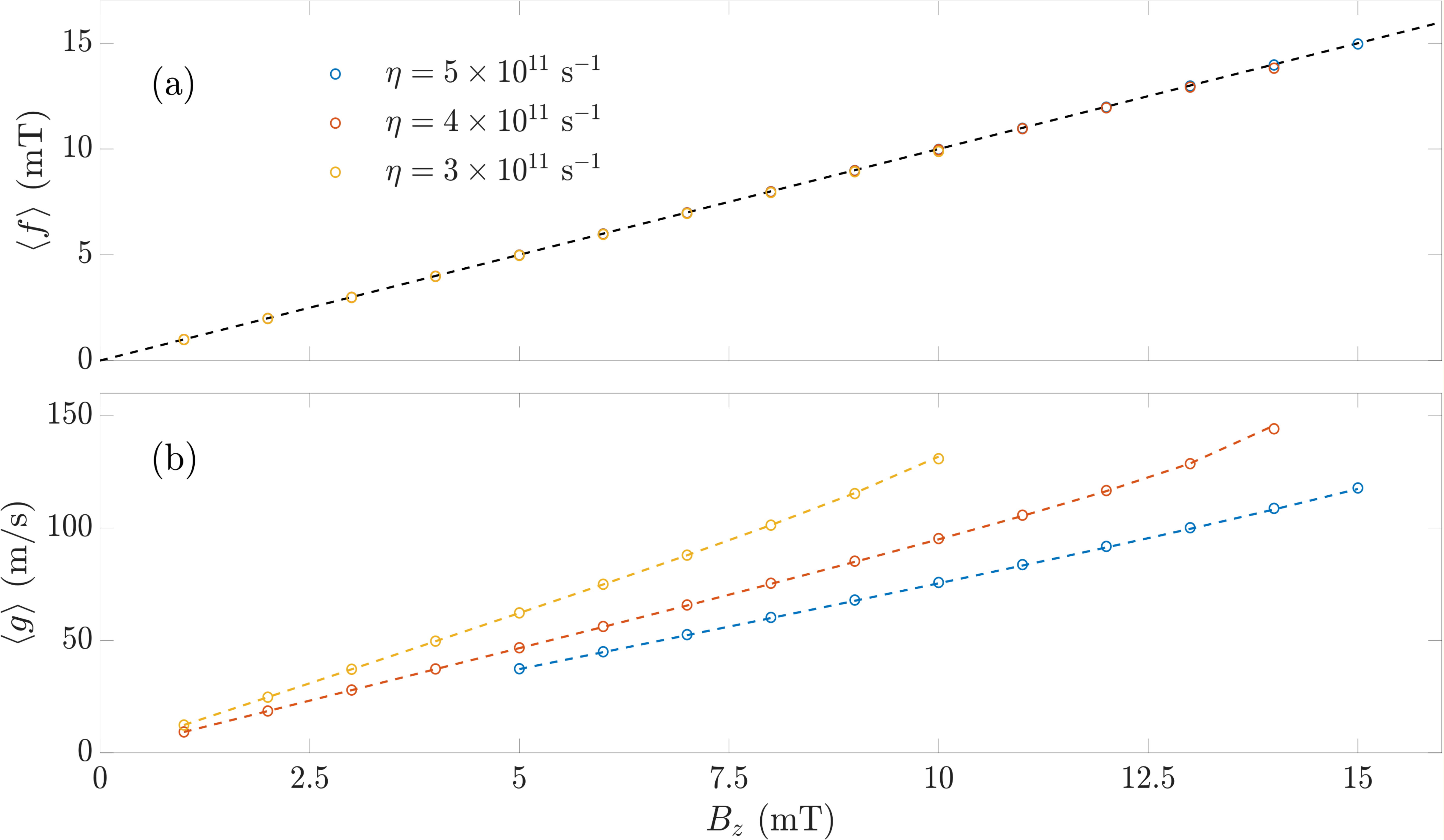}
    \caption{Temporally averaged values of (a) $\langle f \rangle$ and (b) $\langle g \rangle$ as functions of the applied field $B_z$ for different values of elastic damping $\eta$, where $f(t)$ and $g(t)$ represent the right-hand sides of Eqs.~(\ref{eq_f2}) and (\ref{eq_g2}), respectively. Dashed lines correspond to the identity reference curves: (a) $\langle f \rangle = B_z$ and (b) $\langle g \rangle = v_\text{dw}$.}
    \label{fig_en0_fg}
\end{figure}

Figure~\ref{fig_en0_fg}(a) displays the temporal average of the right-hand side of Eq.~(\ref{eq_f2}), denoted as $\langle f \rangle$, evaluated across various applied fields $B_z$ and mechanical damping parameters $\eta$. The theoretical predictions match the applied field $B_z$ with remarkable precision over the entire field range. The robustness of Eq.~(\ref{eq_f2}) sheds key light on how dynamic equilibrium is established in the absence of intrinsic magnetic damping ($\alpha = 0$). Specifically, it reveals the physical origin of the restoring torque required to balance the continuous Larmor precessional torque exerted by the external field $B_z$. As expressed in Eq.~(\ref{eq_f2}), this restoring torque is directly provided by the spatial stress gradients $\langle\sigma_{xx}'\rangle$ and $\langle\sigma_{xy}'\rangle$ centered at the DW core, along with the out-of-plane shear stress component $\langle\sigma_{xz}\rangle$. All three terms act constructively to maintain a stable dynamic steady state.

On the other hand, an expression for the instantaneous DW velocity $v_\text{dw} = \dot{q}$ can be derived from Eq.~(\ref{eq_theta_dot_eta_not_0}) by employing the kinematic relation $\dot{\theta} = -\frac{Q}{\Delta} \sin\theta \, \dot{q}$:

\begin{equation}
    \label{eq_g}
    v_\text{dw} \sin\theta = \frac{\gamma \, Q \, \Delta}{M_s} \left[ \left(K_1 + \frac{3}{2} \lambda_{100} \,\sigma_{xx} \right) \sin(2\varphi) \sin\theta - 3 \lambda_{111} \,\sigma_{xy}\cos (2\varphi) \sin\theta + 3 \lambda_{111} \,\sigma_{xz}\sin\varphi\cos\theta \right].
\end{equation}\noindent 

Unlike Eq.~(\ref{eq_f}), which is strictly valid in the stationary regime, Eq.~(\ref{eq_g}) holds generally across both the steady and precessional propagation regimes. Applying the same procedure as before—multiplying each term by $\sin\theta$ and integrating over $x$—gives the integrated velocity relation:

\begin{equation}
    \label{eq_g2}
    v_\text{dw} = \frac{\gamma \, Q \, \Delta}{M_s} \left[ \left(K_1 + \frac{3}{2} \lambda_{100} \,\langle \sigma_{xx} \rangle \right) \sin(2\varphi) - 3 \lambda_{111} \,\langle \sigma_{xy} \rangle \cos (2\varphi) + 3 \lambda_{111} \, \langle \sigma_{xz} \rangle \sin\varphi \right],
\end{equation}

\noindent where the wall-averaged stress tensor components are defined as

\begin{equation}
\langle \sigma_{xx} \rangle = \frac{\int \sigma_{xx} \sin^2\theta \, \mathrm{d}x}{\int \sin^2\theta \, \mathrm{d}x}, \quad \langle \sigma_{xy} \rangle = \frac{\int \sigma_{xy} \sin^2\theta \, \mathrm{d}x}{\int \sin^2\theta \, \mathrm{d}x}, \quad \langle \sigma_{xz} \rangle = \frac{\int \sigma_{xz} \sin\theta \cos\theta \, \mathrm{d}x}{\int \sin^2\theta \, \mathrm{d}x}.
\end{equation}

Figure~\ref{fig_en0_fg}(b) presents the temporally averaged value $\langle g \rangle$, where $g(t)$ denotes the right-hand side of Eq.~(\ref{eq_g2}), as a function of the applied field $B_z$ across several values of mechanical damping $\eta$. Once again, the theoretical predictions match the numerically simulated velocities with remarkable precision, offering clear physical insight into forward wall propagation. 

\subsection{Discussion}
To understand how mechanical dissipation DW motion, it is instructive to evaluate the relative contributions of the three stress components in Eq.~(\ref{eq_g2}). Both shear stress components, $\sigma_{xy}$ and $\sigma_{xz}$, remain in magnitude small compared to the normal stress component $\sigma_{xx}$ (see Fig.~\ref{fig_en0_sigma}). Consequently, the explicit mechanical contribution to the propulsion of the DW is almost entirely dominated by $\sigma_{xx}$. Because $\sigma_{xx} < 0$ [Fig.~\ref{fig_en0_sigma}(a)] and $\lambda_{100} < 0$, this magnetoelastic coupling term acts constructively alongside the intrinsic hard-axis anisotropy ($K_1$) to generate a strong positive driving force.

Remarkably, unlike in standard Gilbert-damped dynamics ($\alpha \neq 0$), the external driving field $B_z$ does not enter Eq.~(\ref{eq_g2}) explicitly. Instead, its influence is mediated entirely indirectly—first by establishing the stationary tilting angle $\varphi^*$, and second by inducing the localized stress profiles that directly exert the propelling force on the wall.

Furthermore, beyond being the dominant stress contribution, the spatial variations in $\sigma_{xx}$ remain small relative to its static equilibrium value (Fig.~\ref{fig_en0_sigma}). This justifies the constant approximation $\sigma_{xx} \approx \sigma_{xx}^\text{eq} = \frac{1}{2}\lambda_{100}C_{11}$. Within this limit, the domain wall velocity simplifies to

\begin{equation}
    \label{eq_g_approx}
    v_\text{dw} =  \frac{\gamma \, Q \, \Delta}{M_s} \left(K_1 + \frac{3}{4} \lambda_{100}^2 \,C_{11} \right) \sin(2\varphi) \, ,
\end{equation}

\noindent which indicates that the DW motion reverses direction at the pure N\'eel ($\varphi = 0, \pi$) and Bloch ($\varphi = \pi/2, 3\pi/2$) configurations, reaching its peak velocity at $\varphi = \pi/4$. For our system parameters, this theoretical maximum velocity evaluates to $v_\text{dw}^{\text{max}} = \frac{\gamma Q \Delta}{M_s} \left(K_1 + \frac{3}{4} \lambda_{100}^2 C_{11} \right) = 161.71~\text{m/s}$.

\begin{figure}[!htb]
    \centering
   % \includesvg[width=1.0\linewidth]{fig_en0_vvsphi.svg}
     \includegraphics[width=1.0\linewidth]{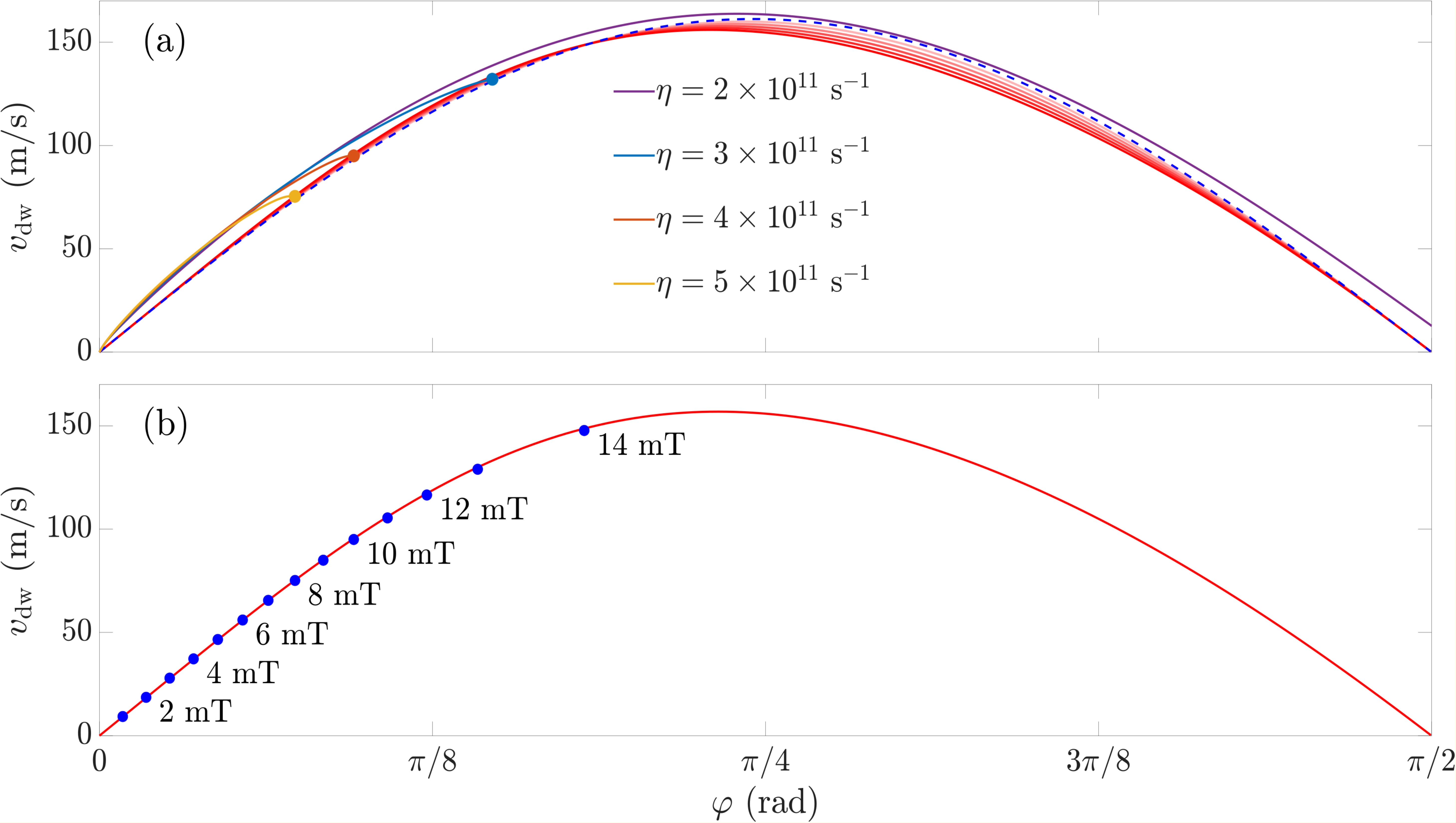}
    \caption{(a) Trajectories in the $v_\text{dw}$--$\varphi$ phase space under an applied field $B_z = 10~\text{mT}$ for different elastic damping values $\eta$. Solid dots highlight the stationary steady state reached in each case. For the lowest damping value ($\eta = 2 \times 10^{11}~\text{s}^{-1}$), no steady state is established, leading to sustained precessional motion (purple curve). Solid red curves depict $v_\text{dw}(\varphi)$ according to Eq.~(\ref{eq_g}) evaluated at the wall center ($\theta = \pi/2$) for increasing $\eta$ (darker red represents higher $\eta$). The dashed blue line corresponds to the analytical approximation in Eq.~(\ref{eq_g_approx}). (b) $v_\text{dw}(\varphi)$ trajectory according to Eq.~(\ref{eq_g}) for $\eta = 4 \times 10^{11}~\text{s}^{-1}$. Steady states for various applied fields $B_z$ are marked with blue dots.}
    \label{fig_en0_vvsphi}
\end{figure}

Figure~\ref{fig_en0_vvsphi}(a) compares the $v_\text{dw}(\varphi)$ response from Eq.~(\ref{eq_g}) evaluated at the wall center ($\theta = \pi/2$) across various elastic damping values $\eta$ (solid curves; color intensity increases with $\eta$) against the analytical approximation of Eq.~(\ref{eq_g_approx}) (dashed blue curve). The approximation captures the overall phase-space trajectory remarkably well across the range of $\eta$ investigated. As $\eta$ increases, both the peak velocity and its corresponding angle decrease slightly; for the maximum damping value, $\eta = 5 \times 10^{11}~\text{s}^{-1}$, we obtain $\varphi^{\text{max}} = 41.25^\circ$ and $v_\text{dw}^{\text{max}} = 156.50~\text{m/s}$. Figure~\ref{fig_en0_vvsphi}(a) also illustrates dynamic trajectories in the $v_\text{dw}$--$\varphi$ phase space under a fixed drive field of $B_z = 10~\text{mT}$. We observe transient behavior where both velocity and tilting angle increase simultaneously until reaching a stationary steady state (marked by solid dots). For the lowest damping case ($\eta = 2 \times 10^{11}~\text{s}^{-1}$), no stationary point is established, and the wall enters a continuous precessional regime (purple trajectory).

Figure~\ref{fig_en0_vvsphi}(b) details the $v_\text{dw}(\varphi)$ curve calculated from Eq.~(\ref{eq_g}) for $\eta = 4 \times 10^{11}~\text{s}^{-1}$ (solid red curve), with stationary states for increasing magnetic fields $B_z$ marked by blue dots. Through the framework of Eqs.~(\ref{eq_f2}) and (\ref{eq_g2}), the origin of the velocity enhancement with driving field becomes clear: to compensate the magnetic field and suppress precession ($\dot{\varphi}=0$), stronger stress amplitudes and steeper stress gradients are required, which according to Eq.~(\ref{eq_sigma_xi}), are achieved for bigger domain wall velocities. Increasing the velocity results in higher equilibrium tilting angles $\varphi^*$.

Remarkably, the steady-state velocity response exhibits a pronounced deviation from linearity as the system approaches Walker breakdown. In standard uncoupled systems with intrinsic magnetic damping ($\alpha \neq 0$), field-dependent wall contraction ($\Delta \propto K_{\text{eff}}^{-1/2}$) produces a subtle sublinear velocity curve near Walker breakdown. In our magnetoelastically coupled system, however, this contraction is overcompensated by the field-induced growth of the stress components. Higher field levels amplify the stress magnitudes, acting both directly as explicit velocity drivers in Eq.~(\ref{eq_g2}) and indirectly by driving the equilibrium angle $\varphi^*$ toward higher values in Eq.~(\ref{eq_f2}). This synergistic mechanism generates a distinct supralinear velocity response prior to breakdown, as clearly evidenced in Figs.~\ref{fig_en0_vvsB}(a) and \ref{fig_en0_vvsphi}(b).

This behavior highlights a central feature of strain-mediated wall dynamics: while the phenomenological progression remains qualitatively reminiscent of magnetically damped systems, the underlying elastodynamic feedback introduces a subtle interplay of direct and indirect force channels that fundamentally reshapes the dynamic phase space.

\begin{acknowledgments}
{We gratefully acknowledge financial support from MICIU/AEI/10.13039/501100011033 and by FEDER, UE, under
Project No. PID2023-150853NB-C31.}
\end{acknowledgments}

\section{Conclusions}
\label{sec_conclusions}

In summary, we have investigated the field-driven dynamics of magnetic domain walls in magnetostrictive nanostripes using a fully coupled micromagnetic-elastodynamic framework alongside a 1D collective coordinates model. By resolving both spin and elastic dynamics self-consistently in the subsonic regime ($v_\text{dw} \ll v_{\ell,t}$), we have demonstrated that mechanical degrees of freedom fundamentally reshape domain wall propulsion and energy dissipation pathways.

Our results demonstrate that mechanical damping ($\eta$) opens a non-magnetic relaxation channel that extracts Zeeman energy released by the advancing wall and converts it into lattice dissipation. This mechanism allows for sustained, field-driven domain wall propagation even in the complete absence of intrinsic magnetic Gilbert damping ($\alpha = 0$). We established a direct quantitative equivalence mapping elastic losses onto an effective magnetoelastic Gilbert parameter $\alpha_\text{me} = 3.11 \times 10^{-13}\,\eta\,\text{(s)}^{-1}$, showing that elastic dissipation acts as an additive, independent drag mechanism alongside standard spin relaxation.

Analytically, our co-moving adiabatic model reveals the underlying mechanism driving this motion: mechanical dissipation breaks the spatial symmetry of the co-moving strain tensor, generating localized stress gradients ($\langle\sigma_{xx}'\rangle, \langle\sigma_{xy}'\rangle$) and out-of-plane shear stresses ($\langle\sigma_{xz}\rangle$) across the wall core. These stress profiles exert direct propelling forces that balance the field-induced precessional torque to establish dynamic steady states. Furthermore, the field-induced amplification of these stress profiles gives rise to a distinct supralinear velocity enhancement near Walker breakdown, highlighting key phase-space differences compared to standard purely magnetic systems.

From a broader perspective, these findings emphasize the necessity of fully coupled, bidirectional models when evaluating nanoscale magnetoelastic dynamics, as unidirectional approximations overlook back-action and lattice-mediated energy relaxation. Beyond domain walls, this strain-mediated dissipation mechanism provides a new design parameter for acoustic spintronics, suggesting that tailored mechanical damping in hybrid piezoelectric or magnetostrictive heterostructures can be exploited to optimize speed, mobility, and energy consumption in domain wall and topological texture devices.

\bibliographystyle{apsrev4-2}
\bibliography{references}

\appendix

\section{General fully-coupled formalism}
\label{sec_general_formalism}

To describe the interconnected magnetic and elastic behavior of the ferromagnet, we employ a self-consistent coupled formalism well documented in the literature \cite{Liang_14,Flauger_26}. The time evolution of the system is governed by the Landau-Lifshitz-Gilbert (LLG) equation for the magnetic degrees of freedom and the elastodynamic wave equation for the mechanical response, given respectively by

\begin{subequations}
\label{eq_coupled_equations}
\begin{align} 
    \frac{\partial\bm{m}}{\partial t} &= -\frac{\gamma}{1+\alpha^2} [\bm{m}\times \bm{B}_\mathrm{eff} + \alpha \, \bm{m}\times (\bm{m}\times \bm{B}_\mathrm{eff})] \label{eq_LLG}  \, , \\ 
    \rho\frac{\partial^2\bm{u}}{\partial t^2} &= \nabla\cdot\bm{\sigma} - \rho\,\eta\frac{\partial\bm{u}}{\partial t} \label{eq_elastic_dynamics}  \, ,
\end{align}
\end{subequations}

\noindent where our state variables are the normalized magnetization vector $\bm{m}$ and the displacement field $\bm{u}$. Here $\gamma$ is the gyromagnetic ratio, $\alpha$ is the dimensionless Gilbert damping parameter, $\rho$ is the mass density of the material and $\eta$ the elastic damping constant. The effective field acting on the magnetization is given by $\bm{B}_\mathrm{eff} = -\frac{1}{M_s} \frac{\delta\mathcal{E}_\text{tot}}{\delta\bm{m}}$, where $\mathcal{E}_\text{tot}$ is the total energy density and $M_s$ is the saturation magnetization. The mechanical behavior is determined by the divergence of the elastic stress tensor, $\bm{\sigma}$, where $(\nabla\cdot\bm{\sigma})_i=\partial_j\sigma_{ij}$ using Einstein summation notation.

The constitutive relation determining the elastic stress tensor $\bm{\sigma}$ from the infinitesimal strain tensor $\bm{\varepsilon}$ is given by

\begin{equation}
\label{eq_constitutive_relation}
    \bm{\sigma} = \bm{C} : \bm{\varepsilon}^\mathrm{el} = \bm{C} : (\bm{\varepsilon} - \bm{\varepsilon}^\mathrm{m}) \, ,
\end{equation}

\noindent where $\bm{C}$ is the fourth-order elastic stiffness tensor and the colon operator (:) denotes a double tensor contraction (in component notation, $\sigma_{ij} = C_{ijkl} \varepsilon_{kl}^\mathrm{el}$, where indices $i, j, k, l \in {x, y, z}$ and the Einstein summation convention is implied over the repeated indices $k$ and $l$). The elastic strain tensor can be decomposed into geometric and magnetostrictive contributions $\bm{\varepsilon}^\mathrm{el} = \bm{\varepsilon} - \bm{\varepsilon}^\mathrm{m}$. The geometric strain tensor is defined as

\begin{subequations}
\label{eq_geometric_strain}
\begin{align} 
    \varepsilon_{ii} &= \frac{\partial u_i}{\partial x_i} \, ,\\ 
    \varepsilon_{ij} &= \frac{1}{2} \left( \frac{\partial u_i}{\partial x_j} + \frac{\partial u_j}{\partial x_i} \right),
\end{align}
\end{subequations}

\noindent while the magnetostrictive strain tensor, assuming cubic crystal symmetry, is expressed as

\begin{subequations}
\label{eq_magnetostrictive_strain}
\begin{align} 
    \varepsilon_{ii}^\mathrm{m} &= \frac{3}{2} \lambda_{100} \left( m_i^2 - \frac{1}{3} \right)  \, ,\\ 
    \varepsilon_{ij}^\mathrm{m} &= \frac{3}{2} \lambda_{111} m_i m_j ,
\end{align}
\end{subequations}

\noindent where $\lambda_{100}$ and $\lambda_{111}$ denote the magnetostriction constants associated with the $\langle 100 \rangle$ and $\langle 111 \rangle$ crystallographic directions, respectively.

The magnetoelastic energy density $\mathcal{E}$ is calculated as

\begin{equation} 
\label{eq_me_energy}
    \mathcal{E}_\text{me} = \frac{1}{2} \bm{\sigma} : \bm{\varepsilon}^\mathrm{el},
\end{equation}

\noindent and, therefore, the effective field contains a magnetoelastic contribution ($\bm{B}_\mathrm{me}$) together with the standard exchange, anisotropy, demagnetizing and Zeeman terms, which is given by

\begin{equation} 
\label{eq_me_field}
    \bm{B}_\text{me} = -\frac{1}{M_s} \frac{\partial\mathcal{E}_\text{me}}{\partial\bm{m}} = \frac{1}{M_s} \bm{\sigma} : \frac{\partial\bm{\varepsilon}^\mathrm{m}}{\partial\bm{m}}.
\end{equation}

In conclusion, our model evaluates the dynamics of the magnetic and elastic degrees of freedom (\ref{eq_coupled_equations}) coupled with each other via the magnetostrictive component in the stress tensor (\ref{eq_constitutive_relation}) and the magnetoelastic contribution to the effective field (\ref{eq_me_field}).

\section{Collective coordinates}
\label{sec_collective_coordinates}

In this Appendix, we summarize the 1D collective coordinates framework used throughout the main text to reduce the micromagnetic dynamics to a set of ordinary differential equations (ODEs) for the DW position $q(t)$ and internal azimuthal angle $\varphi(t)$.

In spherical coordinates (see Fig.~\ref{fig_geom}), the Landau-Lifshitz-Gilbert (LLG) equation can be rephrased as

\begin{subequations}
\label{eq_theta_and_phi_dot}
\begin{align} 
    \dot{\theta} &= \frac{\gamma}{1+\alpha^2} (B_\text{eff}^\phi + \alpha \, B_\text{eff}^\theta) \label{eq_theta_dot} \, , \\
    \sin\theta \, \dot{\phi} &= \frac{\gamma}{1+\alpha^2} (-B_\text{eff}^\theta + \alpha \, B_\text{eff}^\phi) \label{eq_phi_dot} \, ,
\end{align}
\end{subequations}

\noindent where $B_\text{eff}^\theta$ and $B_\text{eff}^\phi$ denote the polar and azimuthal components of the effective field, respectively:

\begin{subequations}
\label{eq_eq_B_eff_theta_and_phi}
\begin{align} 
    B_\text{eff}^\theta &= -\frac{1}{M_s} \frac{\delta \mathcal{E}_\text{tot}}{\delta\theta} \label{eq_B_eff_theta}  \, , \\
    B_\text{eff}^\phi &= -\frac{1}{M_s \sin\theta} \frac{\delta \mathcal{E}_\text{tot}}{\delta\phi}.     \label{eq_B_eff_phi} 
\end{align}
\end{subequations}

Within our 1D approximation and assuming a spatially uniform azimuthal angle along the nanowire length ($\phi(x,t) = \varphi(t)$, $\frac{\partial\phi}{\partial x} = 0$), these effective field components reduce to

\begin{subequations}
\label{eq_eq_B_eff_theta_and_phi_2}
\begin{align}   
    B_\text{eff}^\theta &= \frac{2}{M_s} \left\{ A \frac{\partial^2\theta}{\partial x^2} - [K_0+K_1\sin^2\varphi]\sin\theta\cos\theta \right\}-B_z\sin\theta + B_\text{me}^\theta \label{eq_B_eff_theta_2}  \, , \\
    B_\text{eff}^\varphi &= -\frac{K_1}{M_s} \sin (2\varphi) \sin\theta + B_\text{me}^\varphi \label{eq_B_eff_phi_2} .
\end{align}
\end{subequations}

\noindent where the magnetoelastic contributions ($B_\text{me}^\theta,B_\text{me}^\varphi$) are left unexpanded here, as their explicit functional forms depend on the specific physical approximations evaluated in the main text.

The collective coordinates model assumes a rigid Walker-like polar profile:

\begin{align}     
\label{eq_theta_ansatz}
    \theta(x,t) = 2 \tan^{-1}\left[\exp\left(Q\frac{x-q(t)}{\Delta}\right)\right] \, ,
\end{align}

\noindent where $Q=\pm 1$ represents the topological charge, $q(t)$ is the DW position and $\Delta$ is the DW width. In Cartesian coordinates, the magnetization components corresponding to this ansatz read

\begin{equation}
\label{eq_m}
\begin{split}
    m_x(x,t) &= \text{sech}\left(\frac{x-q(t)}{\Delta}\right) \, \cos\varphi(t) \, , \\
    m_y(x,t) &= \text{sech}\left(\frac{x-q(t)}{\Delta}\right) \, \sin\varphi(t) \, ,\\
    m_z(x,t) &= -Q\tanh\left(\frac{x-q(t)}{\Delta}\right) .
\end{split}
\end{equation}

This ansatz (\ref{eq_theta_ansatz}) verifies $\frac{\partial\theta}{\partial x}=\frac{Q}{\Delta}\sin\theta$ and $\dot{\theta}=-\frac{Q}{\Delta}\sin\theta \, \dot{q}$, which, when substituted in equations (\ref{eq_theta_and_phi_dot}-\ref{eq_eq_B_eff_theta_and_phi_2}) allow us to obtain equations for the time evolution of the DW position and internal angle ($q(t),\varphi(t)$), which, together with the DW width, determine completely the DW dynamics. 

As an example, we consider the case without magnetoelastic coupling ($B_\text{me}^\theta=B_\text{me}^\varphi=0$). Under this condition the bracketed term in (\ref{eq_B_eff_theta_2}) vanishes provided the DW width parameter adapts instantly ($\dot{\Delta}=0$) to its equilibrium value,

\begin{equation}
\label{eq_Delta_no_me}
    \Delta(t) = \left( \frac{A}{K_0 + K_1 \sin^2\varphi(t)} \right)^{1/2}.
\end{equation}

\noindent Substituting the effective field components [Eqs.~(\ref{eq_B_eff_theta_2}) and (\ref{eq_B_eff_phi_2})] into the spherical LLG equations [Eqs.~(\ref{eq_theta_dot}) and (\ref{eq_phi_dot})] yields the ODEs of motion for the DW position and azimuthal angle:

\begin{subequations}
\label{eq_q_and_phi_dot}
\begin{align}   
    \dot{q} &= \frac{\gamma\, Q \Delta \, }{1+\alpha^2} \left[\frac{B_1}{2}\sin(2\varphi) +\alpha \,B_z \right] \label{eq_q_dot}  \, , \\
    \dot{\varphi} &= \frac{\gamma}{1+\alpha^2} \left[ B_z - \alpha\frac{B_1}{2} \sin(2\varphi) \right] \, , \label{eq_varphi_dot} 
\end{align}
\end{subequations}

\noindent where $B_1 = \frac{2K_1}{M_s}$.

\end{document}